\documentclass{aa}
\usepackage{graphicx,natbib,amsmath,amssymb}
\usepackage{txfonts}
\usepackage{color}

\begin{document}

\title{A comparison of preprocessing methods for solar force-free magnetic field extrapolation}
\titlerunning{Preprocessing of solar vector magnetograms}

\author{M. Fuhrmann \inst{1} \and N. Seehafer \inst{1} \and G. Valori \inst{2} \and T. Wiegelmann \inst{3}}

\institute{
Institut f\"ur Physik und Astronomie, Universit\"at Potsdam,
	     Karl-Liebknecht-Str. 24/25, 14476 Potsdam, Germany\\
             \email{fuhrmann@agnld.uni-potsdam.de; seehafer@uni-potsdam.de}
\and
LESIA, Observatoire de Paris, CNRS, UPCM, Universit\'e Paris Diderot, 5 Place de Jules Janssen, 92190 Meudon, France \\
            \email{gherardo.valori@obspm.fr
}
\and
Max-Planck-Institut f\"ur Sonnensystemforschung,
Max-Planck-Str. 2,
37191 Katlenburg-Lindau,
Germany\\
\email{wiegelmann@mps.mpg.de}
}

\date{Received .......... /Accepted ..........}

\abstract
{
 Extrapolations of solar photospheric vector magnetograms into three-dimensional
magnetic fields in the chromosphere and corona are usually done under the assumption
that the fields are force-free. This condition is violated in the photosphere itself
and a thin layer in the lower atmosphere above. The field calculations can be improved by preprocessing the photospheric magnetograms. The intention here is to remove a non-force-free component from the data.
}
{
We compare two preprocessing methods presently in use, namely the methods of
Wiegelmann et al. (2006)  and Fuhrmann et al. (2007).
}
{
The two preprocessing methods were applied to a vector magnetogram
of the recently observed active region NOAA AR 10953. We examine
the changes in the magnetogram effected by the two preprocessing algorithms.
Furthermore, the original magnetogram and the two
preprocessed magnetograms were each used as input data for nonlinear force-free field
extrapolations by means of two different methods,
and we analyze the resulting fields.
}
{
Both preprocessing methods managed to significantly decrease the magnetic forces and
magnetic torques that act through the magnetogram area and that can cause incompatibilities with the assumption of force-freeness in the solution domain. The force and torque decrease is stronger for the Fuhrmann et al. method.
Both methods also reduced the amount of small-scale
irregularities in the observed photospheric field, which can sharply worsen the quality of the solutions.
For the chosen parameter set, the Wiegelmann et al. method led to
greater changes in strong-field areas, leaving weak-field areas mostly unchanged,
and thus providing an approximation of the magnetic field vector in the chromosphere,
while the Fuhrmann et al. method weakly changed the whole magnetogram, thereby better preserving
patterns present in the original magnetogram.
Both preprocessing methods raised the magnetic energy content of the extrapolated fields to values above the minimum energy, corresponding to the potential field.
Also, the fields calculated from the preprocessed magnetograms fulfill the solenoidal condition 
better than those calculated without preprocessing.
}
{
}


\maketitle

\section{Introduction}

In order to model and understand the physical mechanisms underlying the various activity
phenomena that can be observed in the solar atmosphere, like, for instance, spots, faculae, flares, and coronal mass ejections, as well as their mutual connections an interactions, the magnetic field vector throughout the atmosphere must be known.
Unfortunately, reliable magnetic field measurements are still restricted to the level of the photosphere, where the inverse Zeeman effect in Fraunhofer lines is observable.
Progress is very slow here,
due to fundamental difficulties in unambiguously deriving the magnetic
field from polarimetric measurements in chromospheric or coronal spectral lines. Other information sources, like non-burst radio emission or
contrast pictures in selected spectral lines or continuum parts of the solar spectrum,
provide only  order-of-magnitude estimates or limited qualitative information.

As an alternative to measurements in the chromosphere and in the corona, the measured
photospheric field may be extrapolated theoretically into these higher layers
of the atmosphere. Such extrapolations are usually done under the assumption that
the magnetic field $\vec{B}$ is
approximately force-free, i.e. characterized by the equations 
\begin{align}
\nabla\times\vec{B}&=\alpha(\vec{r}) \, \vec{B}\, , \label{force-free}\\
\nabla\cdot\vec{B}&=0 \, , \label{divergenceB}
\end{align}
where $\alpha(\vec{r})$ denotes a scalar function of position $\vec{r}$ which, because of Eq.~(\ref{divergenceB}), is constant along the magnetic field lines.
As an implication of the dominance of the magnetic energy density over all other energy densities \citep[cf.][]{gar01},
this approximation is, except for the times of explosive events,
presumably valid from the upper chromosphere
up to coronal heights of $\sim 1R_{\sun}$ above the photosphere.

The development of extrapolation methods began with calculations of
potential fields, which correspond to the case of $\alpha=0$,
either above limited photospheric areas, in particular active regions \citep{sch64,teutanhag77}, or in spherical shells above the full photosphere \citep{schwilnes69,altnew69}. Extrapolation methods using
linear force-free fields, 
which correspond to the case of a spatially constant non-vanishing $\alpha$, followed as a next step.
Corresponding solutions were  given by \citet{nakraa72}, \citet{chihil77}, \citet{see78}, \citet{ali81}, \citet{sem88};
see also reviews and comparisons  by \citet{see82}, \citet{seesta83}, \citet{gar89}, and \citet{sak89}.

Over the last three decades, extrapolation methods
for nonlinear force-free fields were also developed. Here $\alpha$ is allowed to vary spatially. Unlike the extrapolation methods for current-free and constant-$\alpha$ force-free fields, which require only line-of-sight magnetograms,
the non-constant-$\alpha$ force-free fields are calculated from photospheric vector magnetograms. 
Presently  great efforts are made to improve the methods used here.
They include
(i) direct vertical integration schemes \citep{wuchahag85,wuetal90,cupofmsem90,demcupsem92,sonetal06},
(ii) Grad-Rubin methods \citep{grarub58,sak81,alysee93,amaetal97,amaboumik99,amaboualy06,whe04,whe06},
(iii) boundary-integral methods \citep{yansak97,yansak00,yan03,yanli06,hewan06},
(iv) relaxation methods \citep{yanstuant86,mikmcc94,rou96,valklikep05},
and (v) optimization methods \citep{whesturou00,wie04,wie07}.
Discussions and quantitative tests and comparisons of the nonlinear extrapolation methods are found, e.g., in \citet{aly89}, \citet{sak89}, \citet{mccjiamik97}, \citet{schrietal06}, \citet{metetal08}, \citet{schrietal08}, \citet{wie08}, and \citet{derosetal09}.

Applied to semi-analytical force-free test fields \citep{lowlou90,titdem99},
the nonlinear extrapolation methods show encouraging results \citep{schrietal06,wieetal06,valklifuh07,valetal10}.
 Nevertheless, most of these methods are still lacking a
sound mathematical basis.
The existence of solutions seems to have been proven 
merely for Grad-Rubin methods, and here only
for sufficiently small $|\alpha|$, i.e. close to the potential field
\citep{bin72,kaineuwah00,bouama00}. The Grad-Rubin methods do not exploit the full
information content of photospheric vector magnetograms
(the normal field component $B_n$ and $\alpha$ are used as boundary data, but $\alpha$ is prescribed only
on a part of the boundary where $B_n$ is either positive or negative).

By contrast, the fields calculated by direct vertical integration always satisfy the
conditions given by the photospheric vector magnetograms, and are valid and uniquely
determined solutions to the problem, whenever such solutions exist and are sufficiently smooth. The solutions
are however unstable, or the problem is ill-posed, in the sense that small changes of the
photospheric boundary values can lead to large changes of the field above the
photosphere.

The boundary-integral, relaxation and optimization methods, on the other hand,
impose conditions on all three components of the magnetic vector on the complete boundary of the solution domain, mostly a rectangular cuboid
or a spherical shell above the photosphere,
with some assumtion for the non-photospheric parts of the boundary.
This way the instability of the
solution to boundary-value perturbations is avoided, but inconsistencies in the boundary values lead to only partially force- and divergence-free reconstructions. Therefore, the prescribed boundary values cannot be
arbitrary but have to satisfy consistency conditions. This consistency of the data with the condition of force-freeness
cannot be expected in practice, however, because (i) the field
is not force-free in the photosphere and in the lower chromosphere (possibly with the exception of  regions with particularly strong fields, like sunspots) and (ii)  compatibility with the condition of force-freeness, if it were given for a
real photospheric field, would be destroyed by measurement errors, which are always present, in particular in the transverse
field components (perpendicular to the line of sight of the observer),
e.g. caused by systematic uncertainties
(typically much larger for the transverse than for the longitudinal field
components), or by an inappropriate resolution of the $180^\circ$ ambiguity.

The compatibility of the boundary data with the condition of force-freeness
can be improved by an appropriate preprocessing of the data. Algorithms for such a
preprocessing have been suggested by \citet{wieinhsak06} and \citet{fuhseeval07}. Their strategy is to modify the magnetographic data within certain margins such as to minimize
the total magnetic force and the total magnetic torque on the volume considered.
These
 integral quantities vanish if the magnetic field is force-free, and can be expressed as a function of the boundary values of $\vec{B}$ alone
\citep{mol69,aly84,aly89,low85}. The minimization process
removes a non-force-free component from the data, so that the resulting
boundary values can be more consistently extrapolated into a force-free field.

Another, related objective of the preprocessing is to remove small-scale noise from the data. It has been observed that the convergence of the  evolutionary or iterative sequences calculated using the relaxation or optimization methods
can worsen sharply due such a noise, and the quality of the obtained solutions, that is, the degree to which they
are force-free, is also worsened. This phenomenon is not well understood presently.

One aim of smoothing the photospheric magnetogram is also to
mimic the expansion of the solar magnetic field between photosphere
and chromosphere, where the field is assumed to become force-free.
The idea is that force-free models are only valid above the
chromosphere, but not suitable for modelling the high-beta plasma
between photosphere and chromosphere; see also discussion in \citet{metetal08}.
 In this sense the preprocessed
field may be considered as an approximation to the chromospheric field, which is much smoother and shows less fine structures than the photospheric field.
\citet{jinetal10} compared unprocessed and
preprocessed photospheric field measurements from Hinode with
chromospheric field measurements from Solis and found that the
preprocessed field is indeed a reasonable approximation of the
chromospheric field.

In this paper, we compare the preprocessing methods of \citet{wieinhsak06},
hereafter \emph{ppTW},
and \citet{fuhseeval07}, hereafter \emph{ppMF}. In particular, the two methods are 
applied to a vector magnetogram of a recently observed active region,
NOAA Active Region (AR) 10953, which has been the target in a comparitative study of different extrapolation
methods for nonlinear force-free magnetic fields reported in \citet{derosetal09}.

The remainder of the paper is organized as follows: In Sect.~\ref{sec_prepro} we describe the two preprocessing methods.
Then, in Sect.~\ref{sec_application}, we apply them to
AR 10953: in Sect.~\ref{sec_observations} some observational details of AR 10953
and of the data used are given, and in  Sects.~\ref{sec_effects_magnetogram}
and \ref{sec_effects_field}
the effects of the preprocessing on the magnetogram and on the extrapolated field,
respectively, are analyzed.
In Sect.~\ref{sec_conclusions}, we draw conclusions and discuss our results.

\section{Preprocessing algorithms} \label{sec_prepro}

\subsection{General strategy} \label{sec_prepro_strategy}

The general strategy of \emph{ppTW} and \emph{ppMF}
is to modify the boundary data given by the photospheric vector magnetogram
in order to decrease the total force and the total torque on the volume considered
and to reduce the amount of small-scale noise in the data.
This is done by minimizing a functional $L$ of the boundary values of $\vec{B}$.
$L$ is the sum of a number of subfunctionals
each of which measures one of the quantities to be made small, i.e., the total force, the total torque, and the amount of noise.
The degree of deviation of the modified photospheric fields from the original one during the minimization is controlled either via an additional subfunctional of $L$ that
measures the deviation,
or by allowing the modification of the data only within prescribed domain borders.
The different subfunctionals of $L$ are weighted in order to control their relative importance for the minimization.

\subsubsection{Total magnetic force and total magnetic torque}

Let
\begin{equation} \label{total_force}
\vec{\mathcal{F}}=\int_V \vec{j}\times\vec{B} \, \mathrm{d}V
\end{equation}
denote the total magnetic force and
\begin{equation} \label{total_torque}
\vec{\mathcal{N}}=\int_V \vec{r}\times(\vec{j}\times\vec{B}) \, \mathrm{d}V
\end{equation}
the total magnetic torque on the volume $V$, the solution domain.
In Cartesian coordinates $x$, $y$, $z$, with the $z$ axis pointing upward in the
atmosphere, and with $S$ denoting the magnetogram area in the plane $z=0$,
the condition $\vec{\mathcal{F}}=\vec{0}$ is approximated by
\begin{align}
	\mu_0\, \mathcal{F}_x&=-\int_S{B_x B_z \,\mathrm{d}x\,\mathrm{d}y} =  0 \, ,\label{ff1} \\
	\mu_0\, \mathcal{F}_y&=-\int_S{B_y B_z \,\mathrm{d}x\,\mathrm{d}y}  =  0 \, ,\label{ff2}\\
	2\mu_0 \,\mathcal{F}_z&=\int_S \left( B_x^2 + B_y^2 -B_z^2 \right)\,\mathrm{d}x\,\mathrm{d}y=0 \, ,\label{ff3}
\end{align}
while the condition $\vec{\mathcal{N}}=\vec{0}$ is approximately expressed by 
\begin{align}
2\mu_0\, \mathcal{N}_x&=\int_S y \left( B_x^2 + B_y^2  - B_z^2\right) \,\mathrm{d}x\,\mathrm{d}y=0 \, ,\label{tf1}\\
2\mu_0\, \mathcal{N}_y&=\int_S x \left(- B_x^2 - B_y^2  + B_z^2\right) \,\mathrm{d}x\,\mathrm{d}y=0 \, ,\label{tf2}\\
\mu_0\, \mathcal{N}_z&=\int_S \left(y B_x B_z - xB_yB_z\right) \,\mathrm{d}x\,\mathrm{d}y =0\label{tf3}
\end{align}
\citep{mol69,aly89,fuhseeval07}. 
For expressing the conditions of vanishing force and torque on $V$ exactly,
the integrals  on the right-hand sides of Eqs. (\ref{ff1})--(\ref{tf3})
would have to be extended to integrals over the complete surface $\partial V$ of $V$.
The restriction to integrations over the photospheric magnetogram area $S$
is done under the assumption that
 all relevant magnetic flux is closed on the photosphere, and the field on the rest of the boundary is so weak that its contribution to the surface integrals for $\vec{\mathcal{F}}$ and $\vec{\mathcal{N}}$ is negligible.

\subsubsection{Smoothing} \label{sec_smoothing}

There are several methods for removing small-scale irregularities from a noisy vector magnetogram, hereafter called smoothing \citep{recipes89}. 
We restrict ourselves to the discussion of the two smoothing methods employed by the codes discussed below.

The first method consists in applying the two-dimensional Laplacian, $\Delta_{xy}=\partial^2/\partial x^2+\partial^2/\partial y^2$,
to the photospheric field, $\vec B(x,y,z=0)$, and minimizing the integral of the
square of the obtained function, $\big[\Delta_{xy} \vec B(x,y,z=0)\big]^2$, over the magnetogram area.
$\big[\Delta_{xy} \vec B(x,y,z=0)\big]^2$ measures the roughness of the photospheric boundary data, and
minimizing it corresponds to recducing the deviations of the field values at
given points from the mean values of the field in points around them,
in accordance with the mean-value property of the harmonic functions.
 Since the second-order derivatives measure the curvature of the surfaces $B_x(x,y,z=0)$, $B_y(x,y,z=0)$, and $B_z(x,y,z=0)$, decreasing the modulus of $\Delta_{xy} \vec B(x,y,z=0)$  to zero
on $S$ would remove  all global or local maxima or minima of the three field components
(except for those on the boundary line of the magnetogram), including the physically significant ones, and the total energy content of the field might be strongly reduced.
This kind of smoothing may however well mimic the expansion of the magnetic field
between photosphere and chromosphere.

The second smoothing method considered here is the windowed-median method \citep{recipes89}. Compared to the first method, essentially the normal, arithmetic mean
is replaced by the median. That is, one calculates the median of the field values in
a small window around a given grid point and then minimizes the difference between the field value at the grid point and the median of the window. This way one tries
to avoid a too strong influence of large disturbances that occur only
in a small number of points. Also, the structures in the measured fields may be better conserved even if the quantity minimized here is decreased to zero.
In general, however, decreasing the total force and the total torque on the
one hand and smoothing on the other hand are competing objectives, so that a too
strong smoothing would prevent a sufficient reduction of the total force and the
total torque.

\subsection{Preprocessing to mimic the chromospheric field (\emph{ppTW})}

\citet{wieinhsak06} developed a preprocessing routine in which the ideas
 described above are implemented. In discretized form, the square of the total magnetic force, as approximated by Eqs. (\ref{ff1})--(\ref{ff3}),  is given by
\begin{equation}
	L_1 =  \left( \sum_P{B_x B_z} \right)^2 + \left( \sum_P{B_y B_z} \right)^2 + \left( \sum_P{ \left( B_x^2 + B_y^2 - B_z^2\right)} \right)^2 \, ,\label{L1}
\end{equation}
and the square of the total magnetic torque, as approximated by Eqs. (\ref{tf1})--(\ref{tf3}), by
\begin{align}
	L_2  =&  \left( \sum_P {x \left( B_x^2 + B_y^2 - B_z^2 \right)} \right)^2 + \left( \sum_P {y \left( B_x^2 + B_y^2  - B_z^2 \right)} \right)^2 \nonumber \\
	 &+ \left( \sum_P {y B_x B_z - x B_y B_z} \right)^2 \, \label{L2},
\end{align}
where the summation is over the points $P$ of a photospheric grid
\citep[in the numbering of the different subfunctionals we follow][]{wieinhsak06}.

\citet{wieinhsak06} used a Laplacian-smoothing method (first method in Sect.~\ref{sec_smoothing})
in order to reduce the amount of small-scale irregularities in the magnetogram.
The corresponding smoothing functional is calculated as 
\begin{equation}
 L_4^{(\mathrm{TW})} = \sum_P\left[\left(\Delta_{xy} B_x\right)^2 + \left(\Delta_{xy} B_y\right)^2 + \left(\Delta_{xy} B_z\right)^2\right] \,.\label{eq:L4TW}
\end{equation}

A further subfunctional implemented in \emph{ppTW} measures the difference between the preprocessed magnetogram and the original, observed one,  
\begin{equation}\label{L3TW}
 L_3 = \sum_P\left[\left(B_x - B_x^{(\mathrm{obs})} \right)^2 + \left(B_y - B_y^{(\mathrm{obs})} \right)^2 + \left(B_z - B_z^{(\mathrm{obs})} \right)^2\right] \,,
\end{equation}
where $\vec{B}^{(\mathrm{obs)}}$ corresponds to the observed field values.

The complete minimization functional  is then given as the sum
\begin{equation}
 L^{(\mathrm{TW})} = \mu_1^{(\mathrm{TW})} L_1 + \mu_2^{(\mathrm{TW})} L_2 + \mu_3^{(\mathrm{TW})} L_3 + \mu_4^{(\mathrm{TW})} L_4^{(\mathrm{TW})} \,, \label{eq:LTW}
\end{equation}
where $\mu_1^{(\mathrm{TW})}\dots\mu_4^{(\mathrm{TW})}$ are yet undetermined weighting factors.
For minimizing this functional,
\emph{ppTW} employs a fast and simple Newton-Raphson scheme \citep{recipes89}.

As an extension of the functional given by Eq.~(\ref{eq:LTW}), \citet{wieetal08}
implemented another
``$L_5$'' term, which allows to incorporate direct chromospheric
observations, e.g. H$\alpha$ images, into the preprocessing
scheme. This can improve the approximation of the chromospheric
field, in particular the transverse components, from photospheric
measurements by preprocessing.
The \emph{ppTW} preprocessing scheme has currently been implemented and tested
in spherical geometry by \citet{tadwieinh09}.

\subsection{Pattern-preserving preprocessing (\emph{ppMF})}

In \emph{ppMF}, the total magnetic force and the total magnetic torque are calculated
in the same way as in \emph{ppTW}, namely using Eqs. (\ref{L1}) and (\ref{L2}).
Additionally, the two quantities are made dimensionless by normalizing $L_1$ to
\begin{equation} \label{NL1}
N_{L_1} = \left[ \sum_P  \left(\vec{B}^{(\mathrm{obs})}\right)^2  \right]^2 
\end{equation}
and normalizing $L_2$ to
\begin{align} \label{NL2}
N_{L_2} &= \left[\sum_P \sqrt{x^2+y^2} \left(\vec{B}^{(\mathrm{obs})}\right)^2 \right]^2
\nonumber \\
&=h^2\left[\sum_{k=1}^{N_x}\sum_{l=1}^{N_y}\sqrt{k^2+l^2}
\left(\vec{B}^{(\mathrm{obs})}_{kl}\right)^2 \right]^2 \, ,
\end{align}
with $N_x$ and $N_y$ denoting the numbers of grid points  in the $x$ and $y$ directions ($N_x\cdot N_y=N$), $h$ the grid spacing (assumed to be equal in the  $x$ and $y$ directions) and $\vec{B}^{(\mathrm{obs})}_{kl}$ the elements of the magnetogram
matrix. The normalized quantities $L_1/N_{L_1}$ and $L_2/N_{L_2}$ are independent of the units of length and magnetic field and both are
typically $\sim 1$ for a field that is neither force-free nor torque-free.

For the smoothing, \emph{ppMF} uses a windowed-median technique (second method in
Sect.\ \ref{sec_smoothing}). In discretized form, the smoothing functional is given as
\begin{align}
L_4^{(\mathrm{MF})} = \sum_{i=x,y,z} \sum_P & \Big\{ M_{n}\Big[B_i(x-n\cdot h,y-n\cdot h),\dots \nonumber \\
&\dots,B_i(x+n\cdot h,y+n\cdot h)\Big] - B_i(x,y) \Big\}^2 \, , \label{eq:L4MF}
\end{align}
where $n$ is a positive integer number and $M_{n}$  the median of a rectangular window with $(2n+1)\cdot(2n+1)$ grid points centered about the point $(x,y)$. The values at the boundary of the magnetogram, where the method cannot be applied, are left unchanged; these values are expected to be small. Also, in the practical extrapolations an artificial margin where the field vanishes is in general added to the magnetogram.

$L_4^{(\mathrm{MF})}$ is normalized to
\begin{align} \label{NL4}
N_{L_4^{(\mathrm{MF})}} = \sum_{i=x,y,z} \sum_P & \Big\{ M_{n}\Big[B_i^{(\mathrm{obs})}(x-n\cdot h,y-n\cdot h),\dots\nonumber \\
&\dots, B_i^{(\mathrm{obs})}(x+n\cdot h,y+n\cdot h)\Big] + B_i^{(\mathrm{obs})}(x,y) \Big\}^2 \, ,
\end{align}
so that the normalized quantity $L_4^{(\mathrm{MF})}/N_{L_4^{(\mathrm{MF})}}$ is $\sim 1$ for a field of maximum roughness.

The total functional minimized in \emph{ppMF} is then
\begin{equation}
 L^{(\mathrm{MF})} = \mu_1^{(\mathrm{MF})} \frac{L_1}{N_{L_1}} +  \mu_2^{(\mathrm{MF})} \frac{L_2}{N_{L_2}}
+ \mu_4^{(\mathrm{MF})} \frac{L_4^{(\mathrm{MF})}}{N_{L_4^{(\mathrm{MF})}}} \,,\label{eq:LMF}
\end{equation}
with weighting factors $\mu_1^{(\mathrm{MF})}$, $\mu_2^{(\mathrm{MF})}$, and $\mu_4^{(\mathrm{MF})}$,
and for the minimization the method of simulated annealing is used
\citep[see, e.g.,][]{recipes89}. 

Most important, differently from \emph{ppTW}, \emph{ppMF} does not use the subfunctional $L_3$, which
provides a global measure of the difference between the modified and original
photospheric fields.
Instead, domain borders that must not be overstepped are prescribed locally at each
grid point of the vector magnetogram, different for the line-of-sight and transverse field components. In this way, modifications induced by
the \emph{ppMF} algorithm into the measured field can be consistently limited
to an estimation of the measurement errors.
In the present study (Sect.~\ref{sec_application} below) we work with overall
domain borders of 120 G for the normal field and 250 G for the transverse field,
but the method may be refined by applying domain borders that vary from
point to point.

\section{Application of the two preprocessing routines to AR 10953}
\label{sec_application}

In this section, the two preprocessing routines described in Sect.~\ref{sec_prepro},
\emph{ppTW} and \emph{ppMF},  are applied to a  vector magnetogram of AR 10953,
and the changes in the magnetogram due to the preprocessing are examined.
After that,
the preprocessed magnetograms, as well as the original one, are each extrapolated with
two different methods, namely the magnetofrictional relaxation method of \citet{valklifuh07} and the optimization method of \citet{wie04},
and the resulting fields are analyzed.

AR 10953 has recently been studied using vector magnetograms
by  \citet{okaetal08}, \citet{suetal09},
and \citet{derosetal09}. In particular, the main aim of the study by \citeauthor{derosetal09} was to judge the quality of different extrapolation methods. Here, the attention is focused on the implications of using different preprocessing methods.

\subsection{NOAA AR 10953 and data used}
\label{sec_observations}

NOAA AR 10953 was a moderately flare-active region with a basically bipolar magnetic structure.
The vector magnetogram employed here is identical to the one used in the study of \citet{derosetal09}.
It was derived from the Stokes profiles 
of the two magnetically sensitive Fe~I lines at
$6301.5\,\mbox{\AA}$  and $6302.5\,\mbox{\AA}$
measured by
the Spectro-Polarimeter (SP) instrument of the Solar Optical Telescope
 \citep[SOT; ][]{tsuetal08} onboard the \emph{Hinode} satellite \citep{kosetal07}.
The corresponding scan of the active region began at 22:30 UT on 2007 April 30 and took about 30 min. The angular resolution in the East-West and North-South directions was $0.32''$.
The $180^\circ$ ambiguity in the transverse
field was removed by means of the AZAM utility \citep{litetal95,metetal06}.
More details on the preparation of the vector magnetogram may be found in \citet{schrietal08} and
\citet{derosetal09}, and  references given in these papers.

\begin{figure*}
 \centering
 \includegraphics[width=14.0cm]{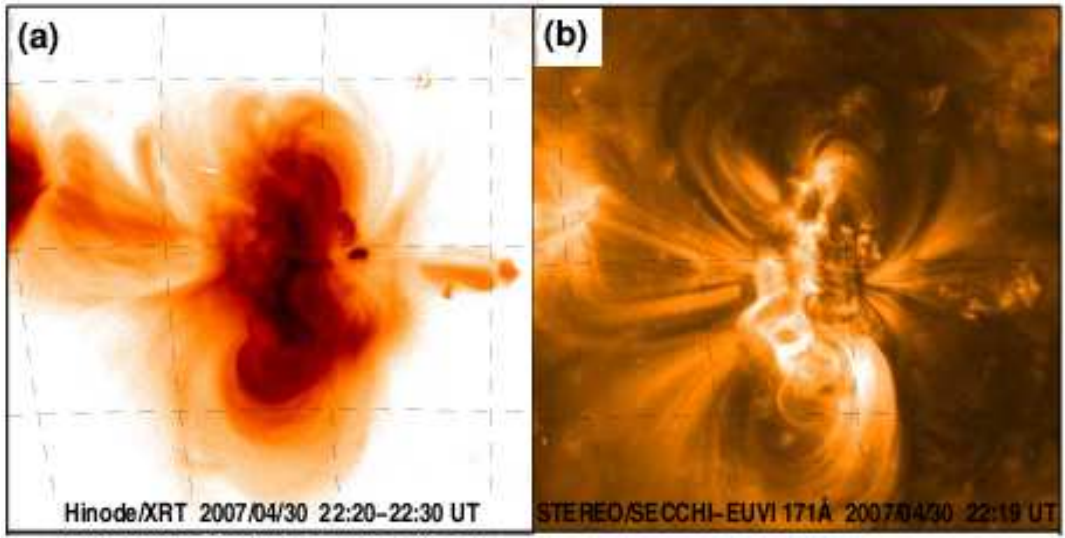}
 \caption{Two coaligned images of AR 10953 (with the same 10 degree gridlines drawn on
both images for reference). (a) Time-averaged and logarithmically scaled \emph{Hinode}/XRT soft X-ray image and (b) STEREO-A/SECCHI-EUVI 171{\AA} image.  }
 \label{fig:derosa}
\end{figure*}

The area of the \emph{Hinode}/SOT-SP scan  covers the central part of AR 10953, which was dominated
by a leading spot of negative polarity.
The active region
 was flare-quiet above the C1.0 level until about two days after the scan.
 Images from the \emph{Hinode} X-Ray Telescope
\citep[XRT; ][]{goletal07,kanetal08} taken around the time of the scan show
  bright loops that probably make visible bundles of coronal magnetic field lines,
see Fig.~\ref{fig:derosa}(a).
Similar bundles of field lines are apparently seen in images from the
Extreme Ultraviolet Imagers (EUVI),  which are parts of the
Sun Earth Connection Coronal and Heliospheric Investigation (SECCHI)  telescope packages
\citep{howetal08}
 onboard the two \emph{Solar Terrestrial
Relations Observatory (STEREO)} spacecraft, see Fig.~\ref{fig:derosa}(b).

A significant part of the area of the vector magnetogram seems to be magnetically connected to regions outside this area. This is, for instance, also indicated by three-dimensional trajectories of loops as  stereoscopically determined \citep{aschwetal08}
from STEREO/SECCHI-EUVI images
 \citep[not shown here, see Fig.~1 in][]{derosetal09}.

In the study of \citet{derosetal09},
the vector magnetogram was therefore embedded in a larger line-of-sight magnetogram with a lower spatial resolution,
observed by the Michelson Doppler Imager (MDI) instrument \citep{scheretal95} onboard the Solar and Heliospheric Observatory (SOHO) spacecraft. 
The  total magnetogram used comprises, in a tangential plane to the photosphere,
$320\times 320$ pixel$^2$, with
$1\,\mathrm{pixel}=580\,\mathrm{km}$,
so that altogether an area of 
$185.6\,\mathrm{Mm}\times 185.6\,\mathrm{Mm}$
 is covered,
whereas the measured full magnetic vector is available
for a subarea of $100\,\mathrm{Mm}\times115\,\mathrm{Mm}$. 
The photospheric $320\times 320$ pixel$^2$ region in turn is part of
an encompassing $512\times 512$ pixel$^2$ region
for which a line-of-sight magnetogram was available from the SOHO/SDI measurements,
 and from
this $512\times 512$ pixel$^2$ line-of sight magnetogram the potential field in a
$512\times 512\times 512$ pixel$^3$ cube above the photosphere was calculated
using a Green's function method described in \citet{metetal08}
(with the normal field component set equal to zero at the side and top faces
of the cube).
 In the study of \citet{derosetal09}, the solution domain
for the nonlinear force-free fields was then a cuboid with a height of 256 pixels above the
$320\times 320$ pixel$^2$ magnetogram, with the boundary conditions at the side and top faces, for the methods that need such conditions, given by the
potential field calculation for the encompassing $512\times 512\times 512$ pixel$^3$ cube, and assuming the photospheric field outside the measured vector
magnetogram to be vertical, i.e., setting the horizontal field components to zero there.

Here, the preprocessing algorithms are applied to the enlarged, $320\times 320$ pixel$^2$
 vector magnetogram. The subsequent analysis of the
effects of the preprocessing on the photospheric field is only in part done for this enlarged magnetogram and otherwise restricted to
the smaller photospheric region where the magnetic vector was measured.

\subsection{Effects of the preprocessing routines on the vector magnetogram}
\label{sec_effects_magnetogram}

In this subsection, the preprocessed vector magnetograms are analysed with respect to the differences they show compared to the observed magnetogram and to each other.
Namely, we try to assess the smoothness of the magnetograms and how well they satisfy the conditions
of force-freeness and torque-freeness. Furthermore, the function $\alpha(\vec{r})$ defined by Eq.\ (\ref{force-free}),
whose values in the level of the
photosphere can be calculated from the magnetogram, is considered. This function contains
much information on the topology of the magnetic field lines. In particular, the field
lines lie in the surfaces $\alpha=\mathrm{constant}$ and, thus, $\alpha$ must have equal
values at any two photospheric points that are connected by a field line above the
photosphere. Finally, the deviations from the original magnetogram produced by the two preprocessing
methods are compared.

The minimization functionals, given by Eqs. (\ref{eq:LTW}) and (\ref{eq:LMF}),
depend on the weighting factors $\mu_i^{(\mathrm{TW})}$
and $\mu_i^{(\mathrm{MF})}$, respectively, which are free parameters (actually only
their ratios count, so that for each of the procedures one of the factors can be set equal to unity). The optimal choice of the weighting factors poses a special problem of the
preprocessing. As mentioned in Sect.~\ref{sec_smoothing}, minimizing the total magnetic force and the total magnetic torque on the one hand and the noise level on the other
hand are competing objectives.
If, for instance, the weight of smoothing in the minimization
functional is set to zero, the subfunctionals measuring the forces and torques
are generally found to be minimized very fast to very small values. If smoothing
is included, however, the reduction of the forces and torques is impeded.
On the other hand, a strong weighting of the forces and torques inhibits
the smoothing \citep[cf. consideration of this problem in][]{fuhseeval07}.

For the results presented in the following, the
weighting factors are chosen such that the competing minimization objectives appear
properly balanced. The weighting factors for \emph{ppTW} are chosen as
$\mu_1^{(\mathrm{TW})}=\mu_2^{(\mathrm{TW})}=1.0$,
$\mu_3^{(\mathrm{TW})}=0.001$,
 $\mu_4^{(\mathrm{TW})}=0.01$,
and those for \emph{ppMF} as
$\mu_1^{(\mathrm{MF})}=\mu_2^{(\mathrm{MF})}=\mu_4^{(\mathrm{MF})}=1.0$.
\emph{ppTW} uses a normalization of the magnetic field to
the mean value of $\left|\vec{B}^{(\mathrm{obs})}\right|$ over the magnetogram
and of lengths to the edge length of the magnetogram; so the subfunctionals of $L^{\mathrm{(TW)}}$ and
their weighting factors  are dimensionless but, unlike the corresponding quantities
in \emph{ppMF}, not independent of the chosen units of magnetic field and length
(thus the parameters cannot be compared directly).
Furthermore, \emph{ppMF} is carried out with a window size of $n=1$ for the smoothing
and with domain borders of $120\,\mathrm{G}$ for the normal field component
and of $250\,\mathrm{G}$ for the transverse field components
during the minimization.

\begin{figure*}
 \centering
\includegraphics[width=0.3\textwidth]{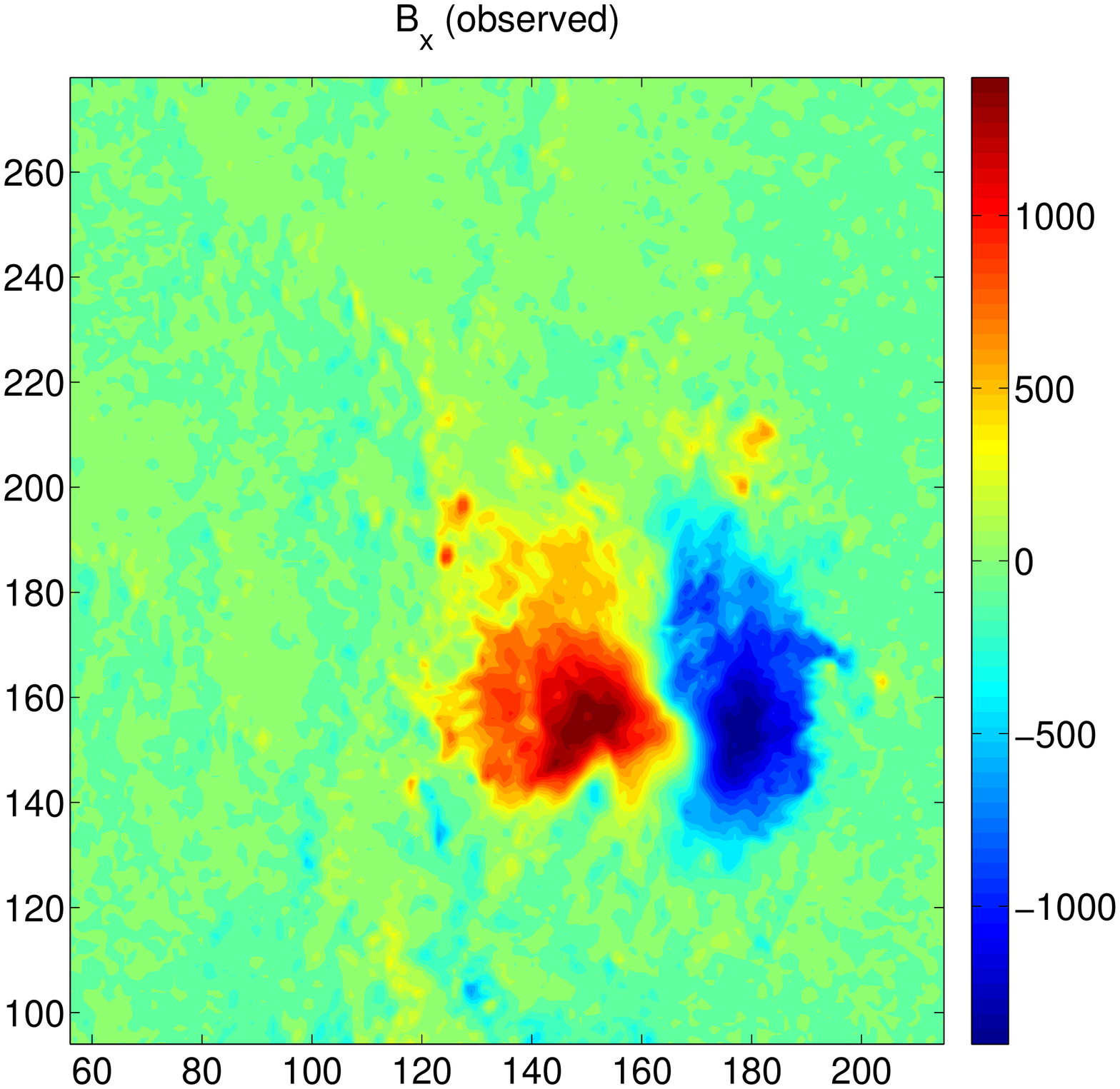}
\includegraphics[width=0.3\textwidth]{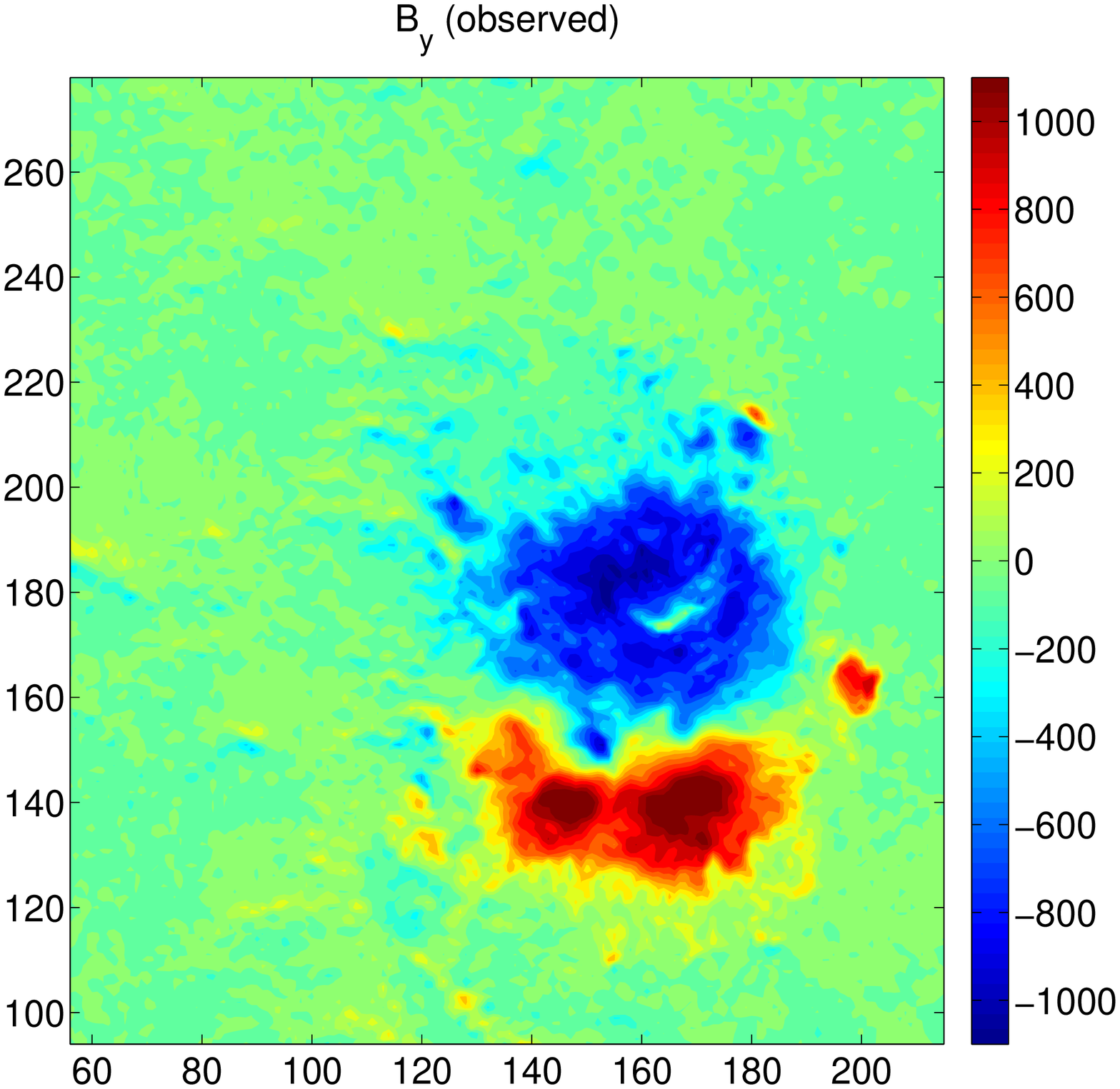}
\includegraphics[width=0.3\textwidth]{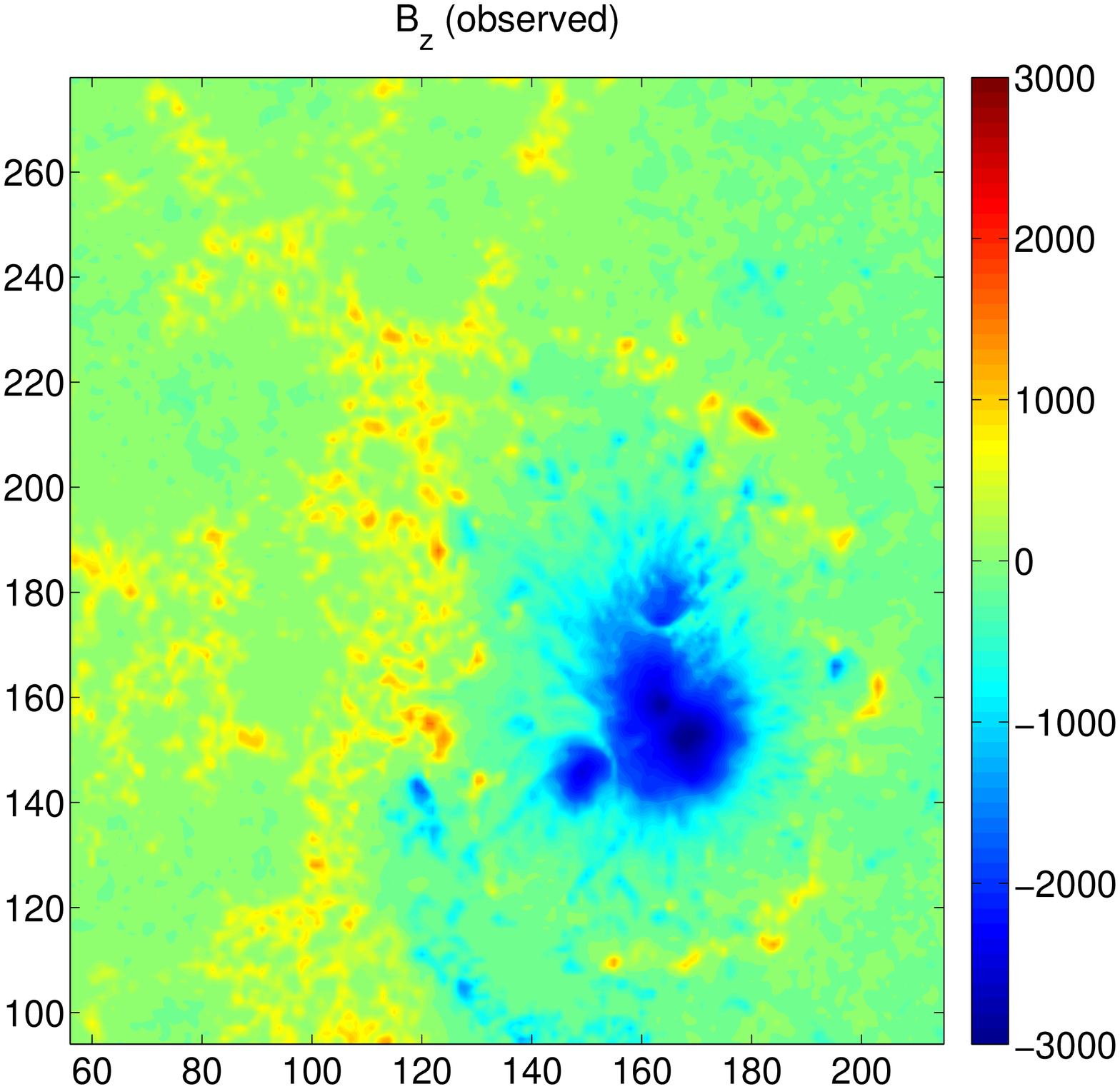} \\
 \includegraphics[width=0.3\textwidth]{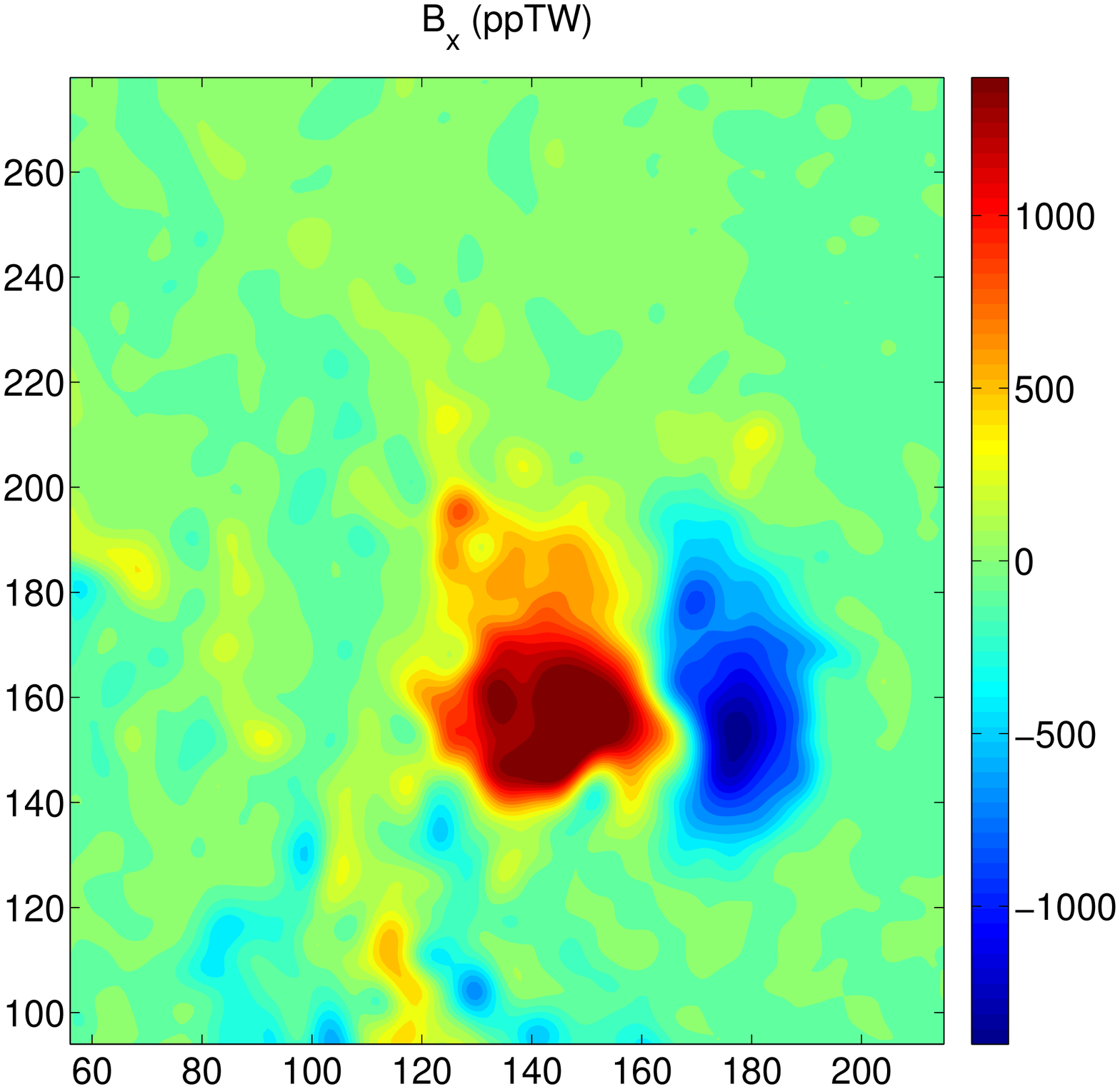}
\includegraphics[width=0.3\textwidth]{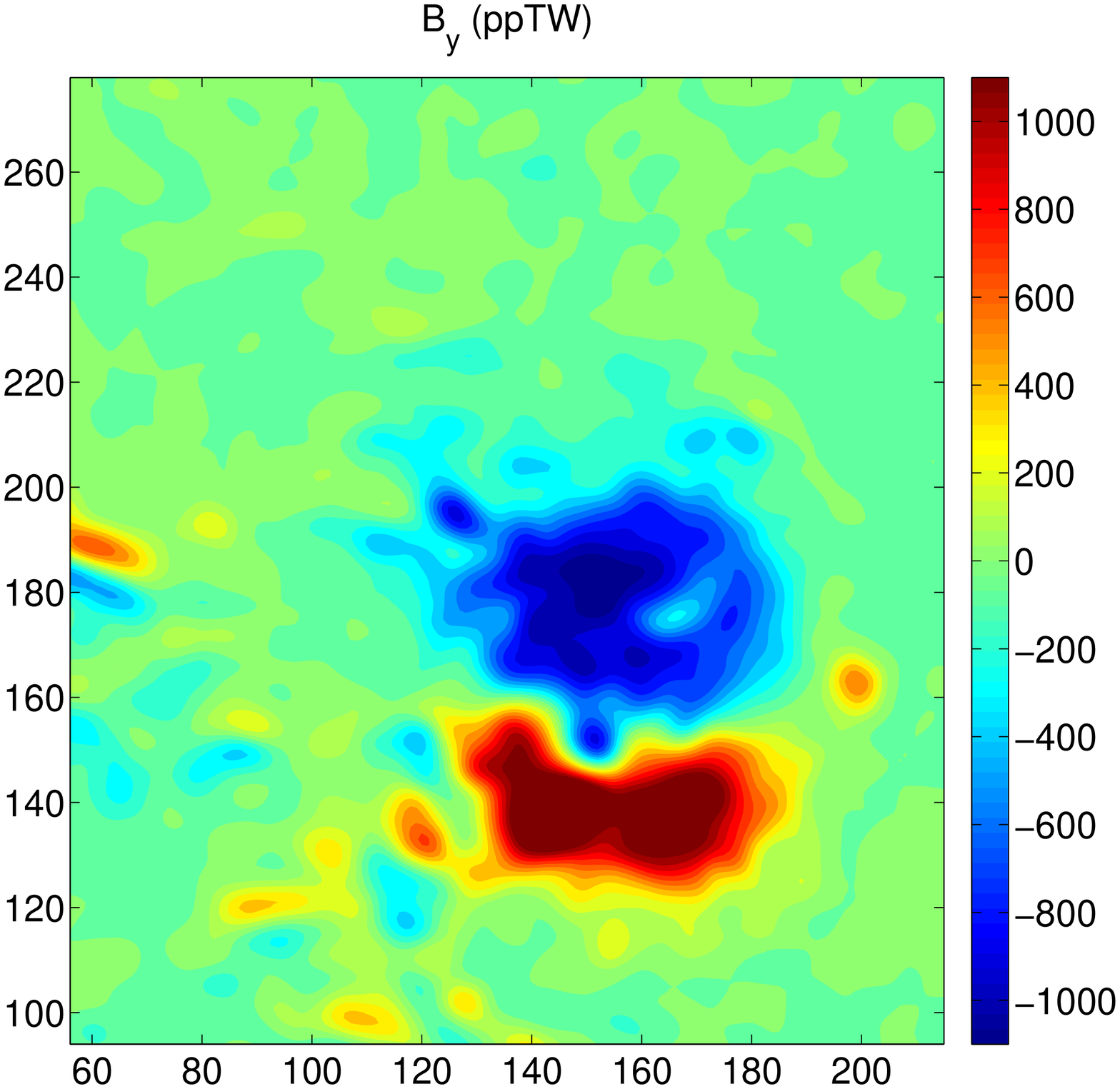}
\includegraphics[width=0.3\textwidth]{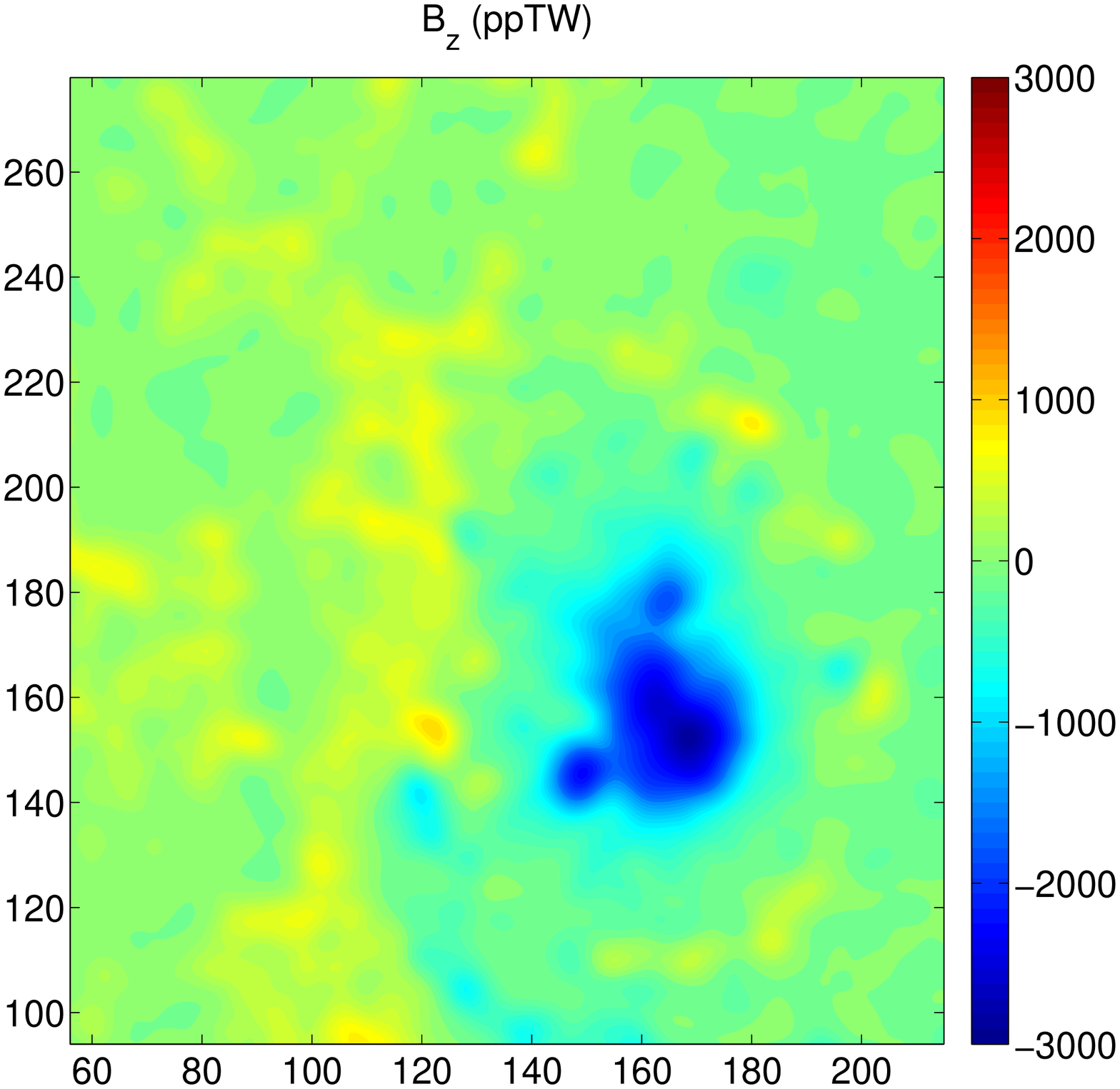} \\
\includegraphics[width=0.3\textwidth]{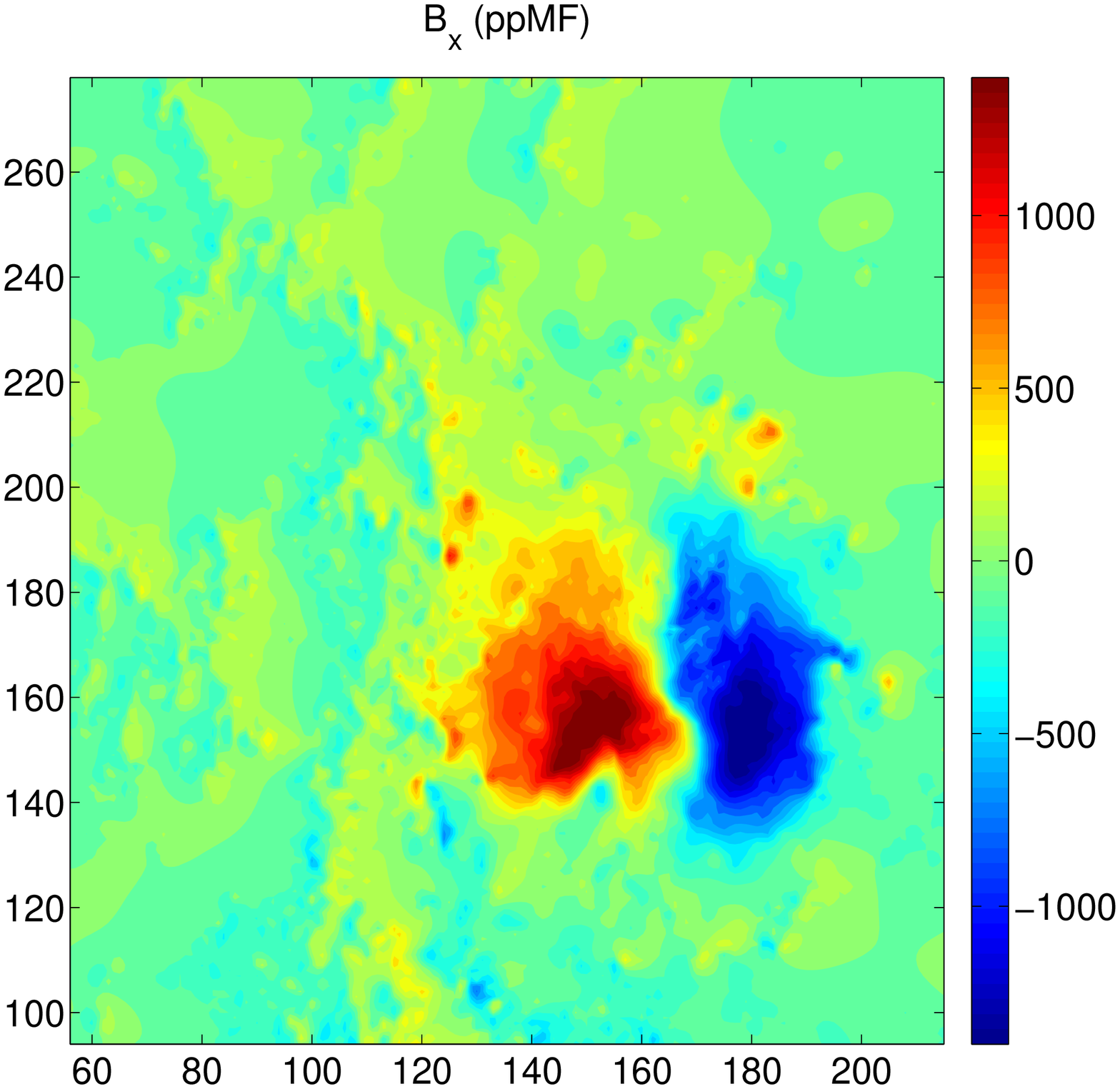}
\includegraphics[width=0.3\textwidth]{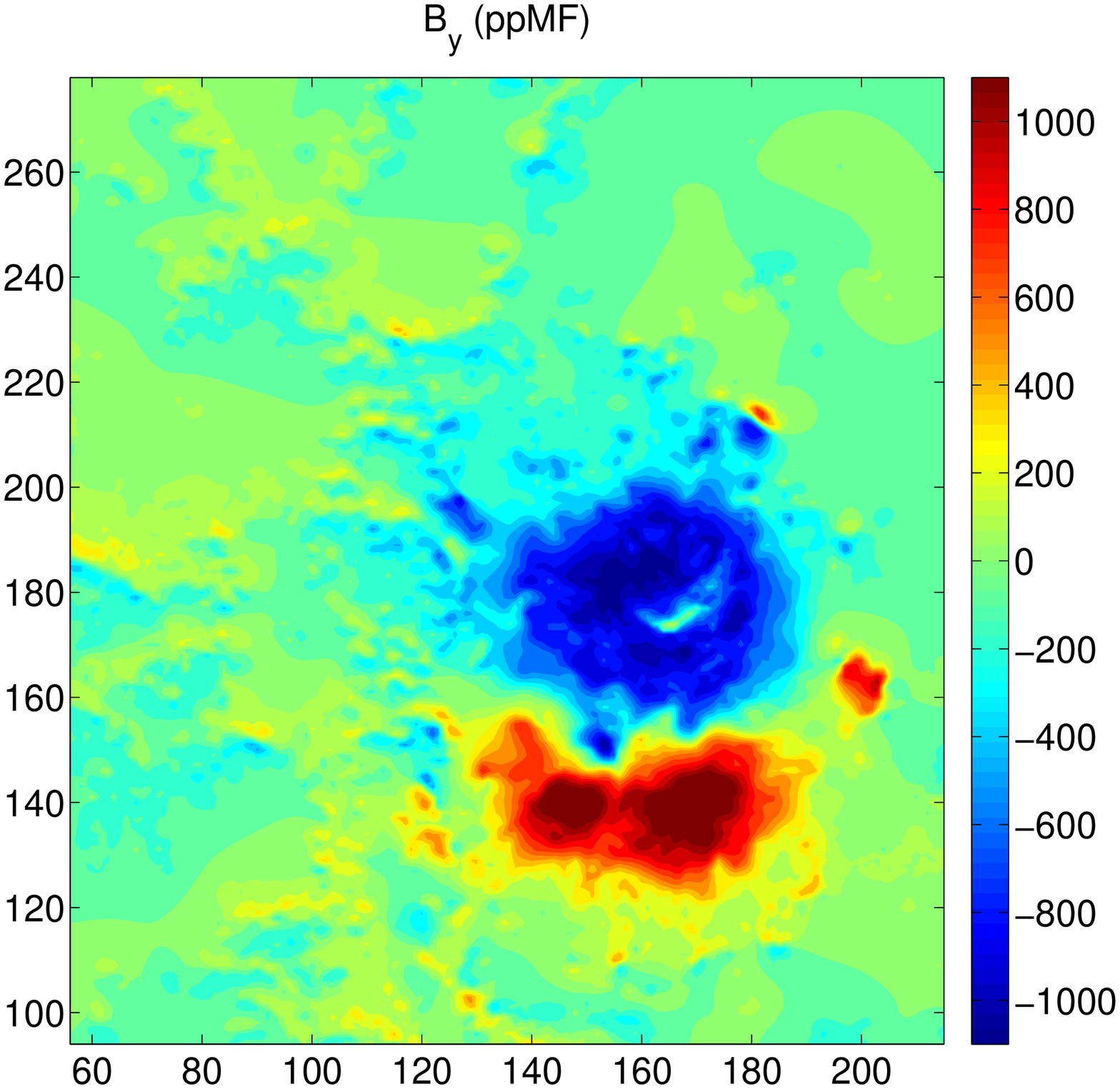}
\includegraphics[width=0.3\textwidth]{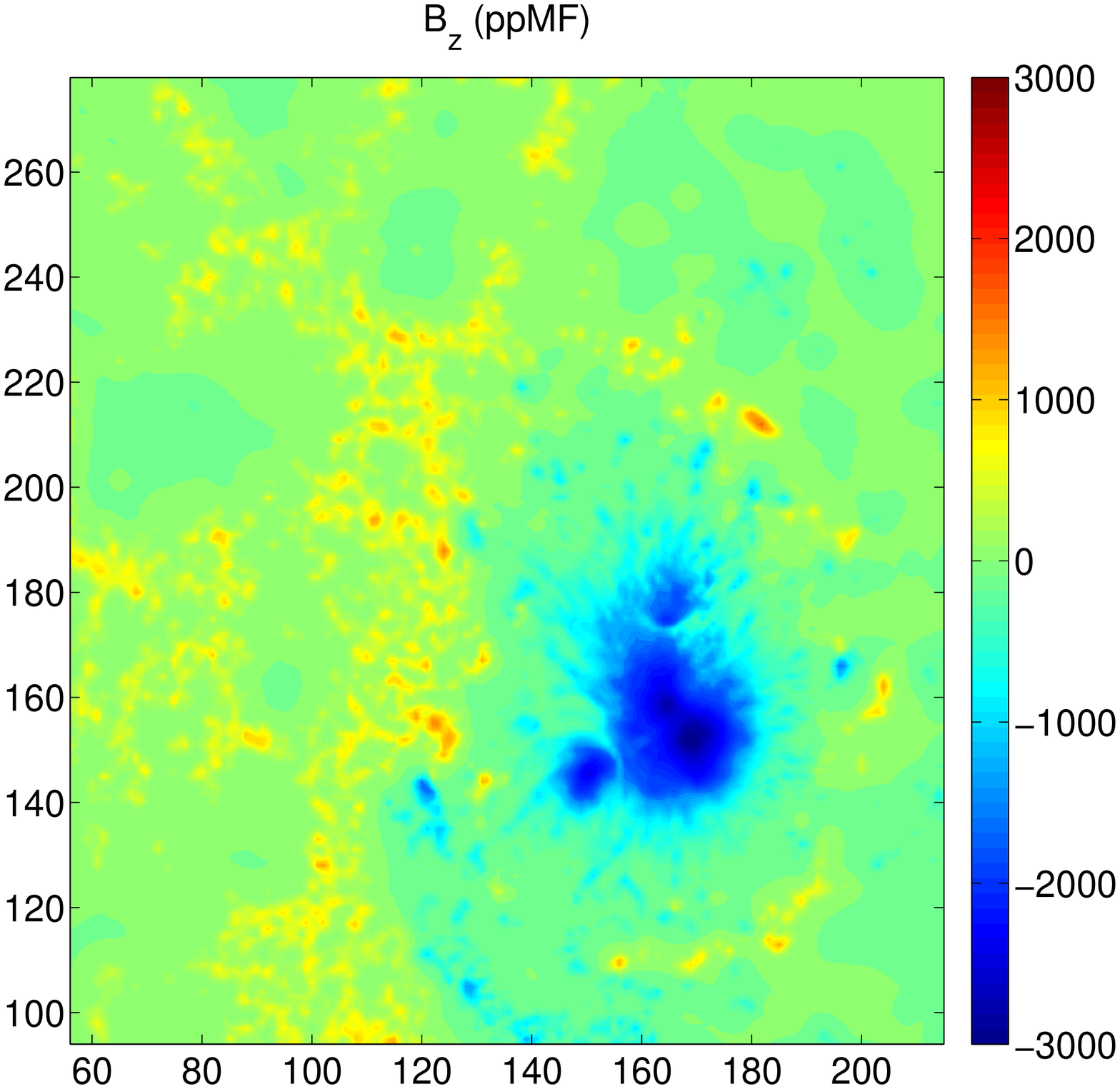}
 
 \caption{Original and preprocessed vector magnetograms of AR 10953. \emph{From left to right:} $B_x$, $B_y$, and $B_z$. \emph{From top to bottom:} Unpreprocessed, \emph{ppTW}, and \emph{ppMF}.
The magnetic field is measured in G and the length unit is pixel ($580\,\mathrm{km}$).}
 \label{fig:mgm10953}
\end{figure*}
\begin{figure*}
 \centering
\includegraphics[width=0.3\textwidth]{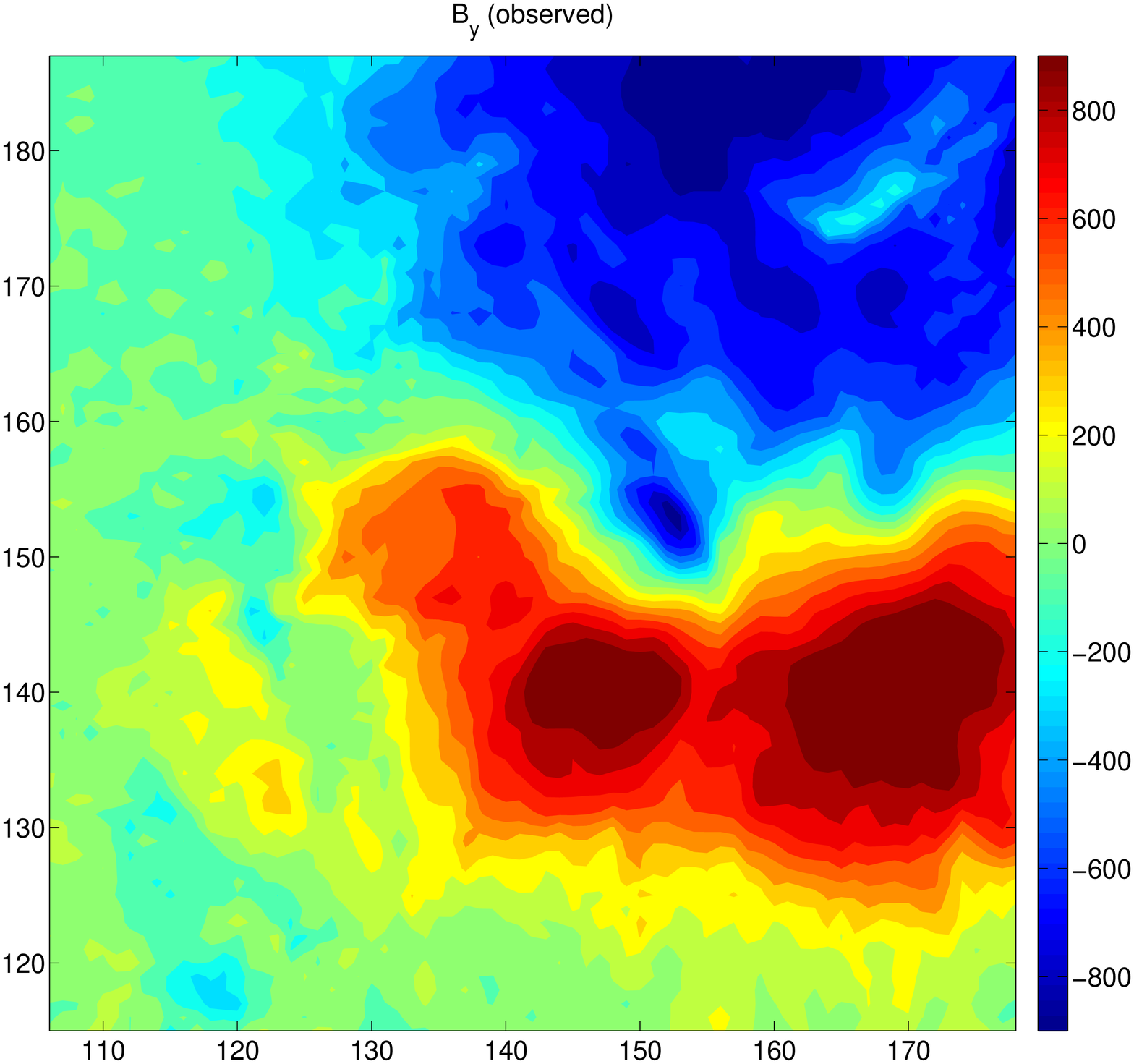}
 \includegraphics[width=0.3\textwidth]{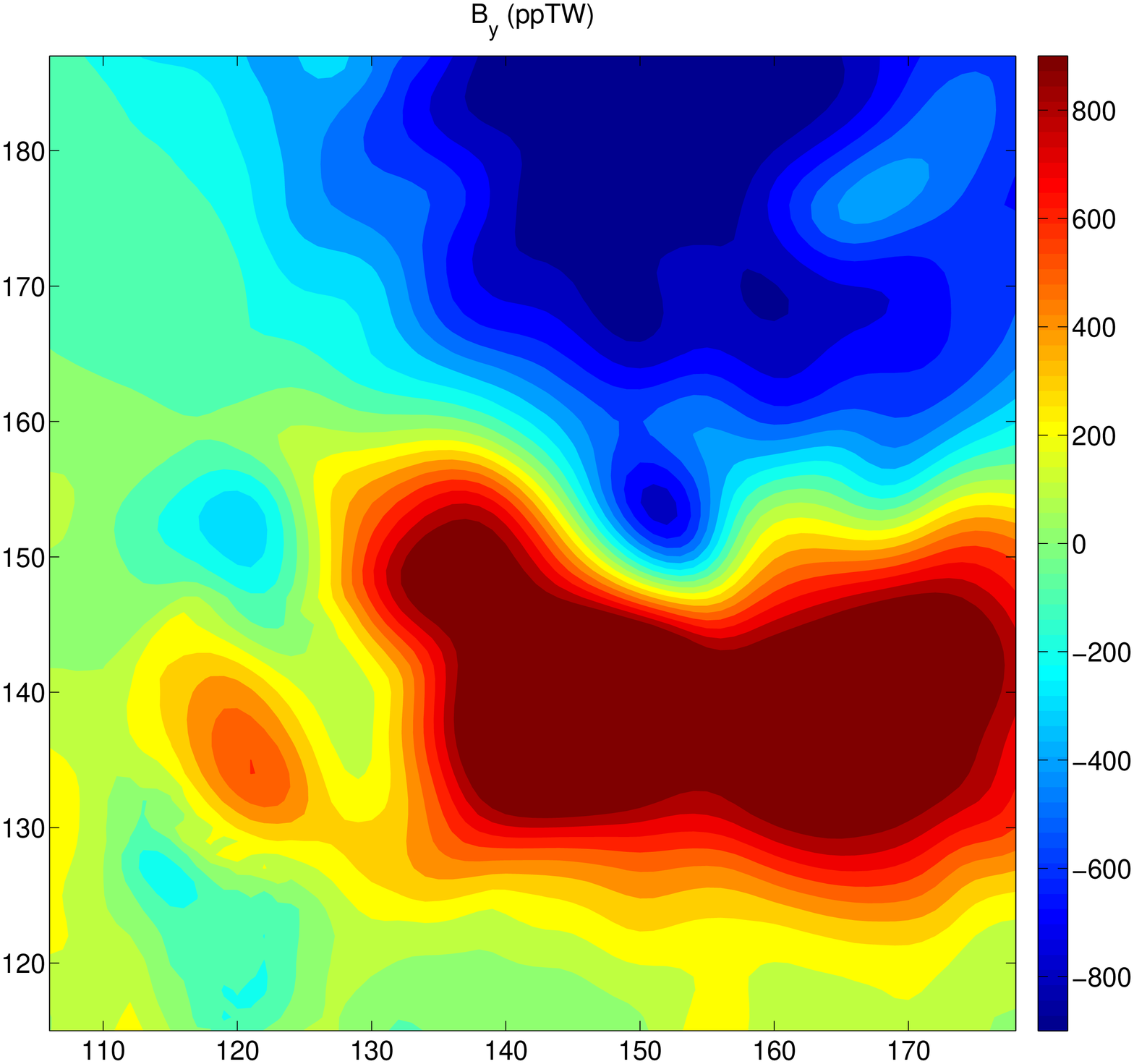}
 \includegraphics[width=0.3\textwidth]{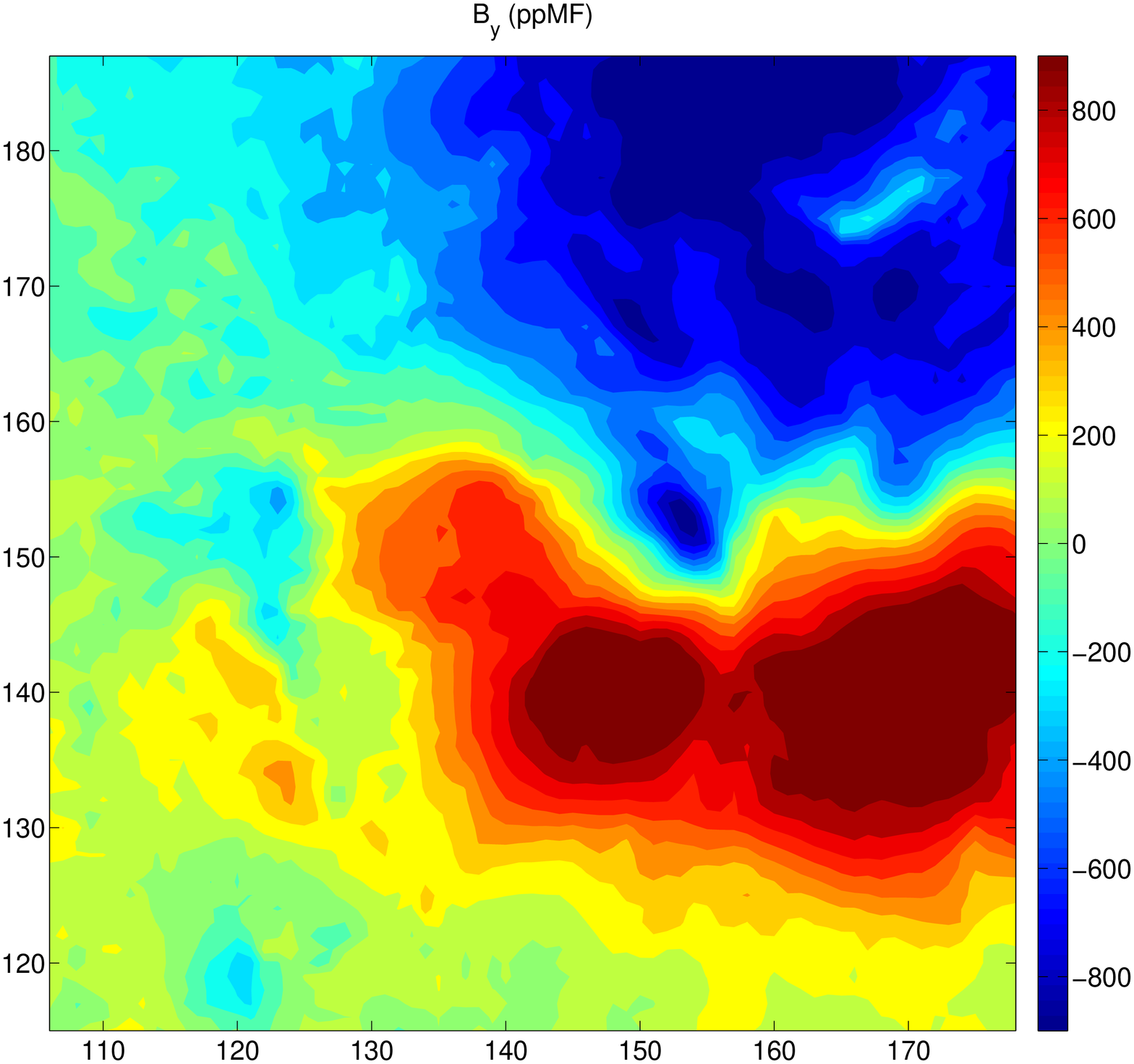}
\caption{$B_y$ in a magnified cutout of the vector magnetogram of AR10953. \emph{From left to right:} Unpreprocessed, \emph{ppTW}, and \emph{ppMF}.
The magnetic field is measured in G and the length unit is pixel ($580\,\mathrm{km}$).}
 \label{fig:bydet}
\end{figure*}

Fig.~\ref{fig:mgm10953} shows outcomes of preprocessing the measured vector magnetogram by the two methods. Obviously,
\emph{ppTW} leads to a markedly smoother vector magnetogram than \emph{ppMF}. Small-scale irregularities are
largely removed by \emph{ppTW}. The Laplacian smoothing used by this method resembles a diffusion process.
Correspondingly, originally separated maxima or minima may be fused, as is seen for the $B_y$ component in
Fig.~\ref{fig:bydet},
where a magnification of the central part of the magnetogram region is shown.
By contrast, though also smoothing the magnetogram, \emph{ppMF} manages
 to stay close to the observed structures. The reason for the different smoothing levels seems to be rooted in the different smoothing schemes. The subfunctional $L_4^{\mathrm{(MF)}}$, defined by Eq.~(\ref{eq:L4MF}), decreases very fast during the preprocessing, without changing the magnetogram very much. Experiments for \emph{ppMF} with a Laplacian smoothing led to much smoother magnetograms, but the decrease rate of the smoothing subfunctional, $L_4^{\mathrm{(TW)}}$,
was smaller than that of the correponding subfunctional, $L_4^{\mathrm{(MF)}}$, when applying the windowed-median smoothing.

A quantitative comparison of the two preprocessing routines is given in Table~\ref{tab:pp}.
\begin{table}
\caption{Quantitative  comparisons of the original and preprocessed vector magnetograms.$^{\mathrm{a}}$}
\begin{center}
\begin{tabular}{cccc} 
 \hline\hline
 & Unpreprocessed & \emph{ppTW} & \emph{ppMF}\\
\hline
Total magnetic force & 6.71 $\cdot 10^{-2}$ & 2.20 $\cdot 10^{-4}$ & 1.48 $\cdot 10^{-6}$ \\
Total magnetic torque & 5.50 $\cdot 10^{-2}$  & 9.83 $\cdot 10^{-5}$ & 5.83 $\cdot 10^{-5}$ \\
Net magnetic flux $M_f$ & -0.13 & -0.13 & -0.15  \\
max $|\alpha(x,y,z=0)|$ &2.73 &1.22 & 2.00 \\
$\left< \vert \alpha(x,y,z=0) \vert \right>$ & 0.057 & 0.055 & 0.062 \\
\hline
\end{tabular}
\label{tab:pp}
\end{center}
$^{\mathrm{a}}$
The total magnetic force, the total magnetic torque, and the net magnetic flux are given in normalized, dimensionless forms (cf. text), and $\alpha$ is
measured in pixel$^{-1}$ ($1\,\mathrm{pixel}=580\,\mathrm{km}$). 
The total magnetic force and the total magnetic torque are calculated for the enlarged,
$320\times 320$ pixel$^2$ magnetogram, the other quantities
for the smaller area where the magnetic vector was measured.
\end{table} 
The values of the total magnetic force and and the total magnetic torque shown there were calculated according to
Eqs.\  (\ref{L1}) and (\ref{L2}) and subsequently normalized to the quantities given by Eqs. (\ref{NL1}) and (\ref{NL2}), respectively. 

\emph{ppTW} was able to decrease the total force and the total torque
 by two orders of magnitude. The measured total force after application of \emph{ppMF} is even decreased by four orders of magnitude compared to the original total force, and the the measured total torque after application
of  \emph{ppMF} is roughly by a factor of two smaller than that after application of \emph{ppTW}.

\begin{figure}
 \centering
 \includegraphics[width=0.45\textwidth]{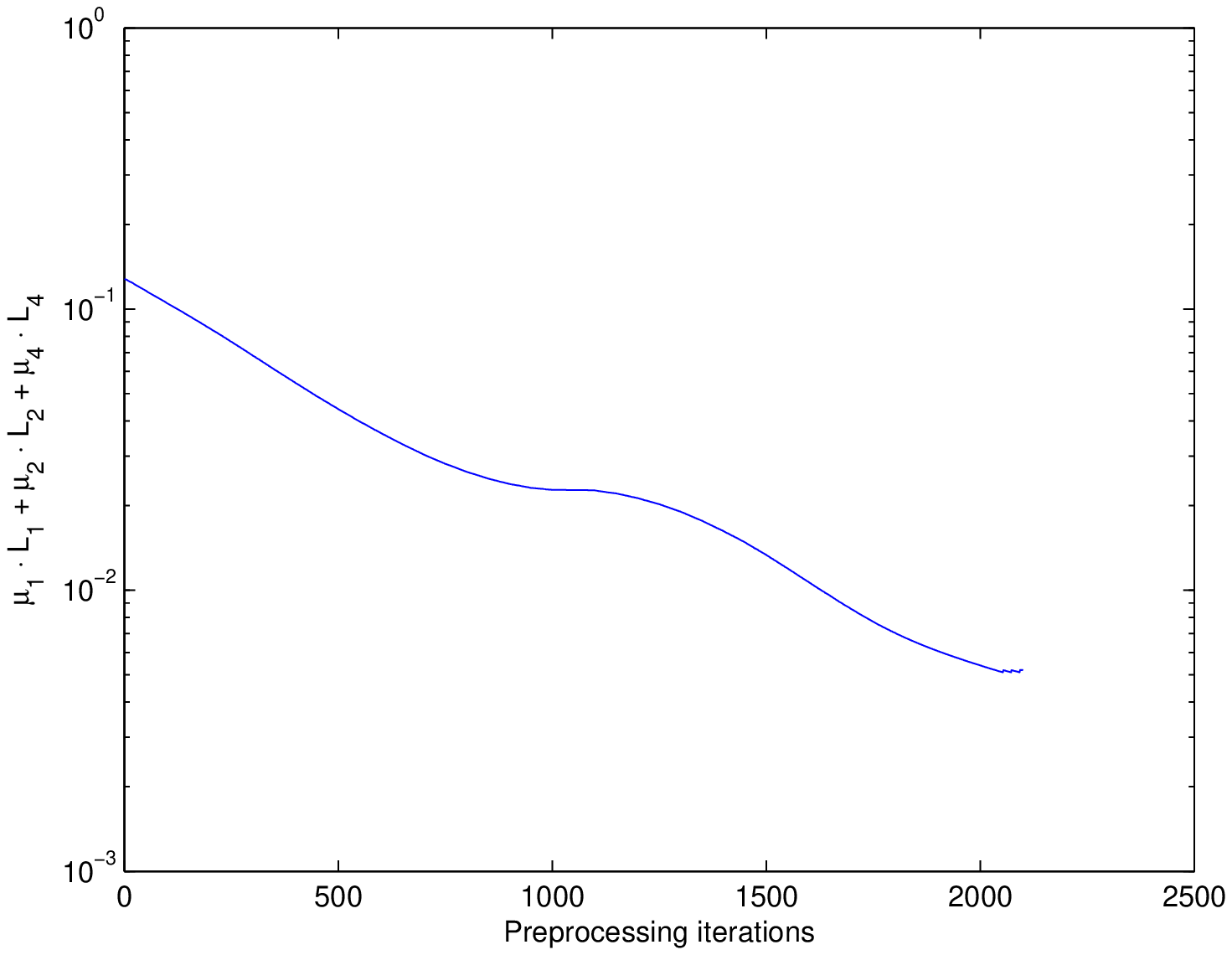}
 \includegraphics[width=0.45\textwidth]{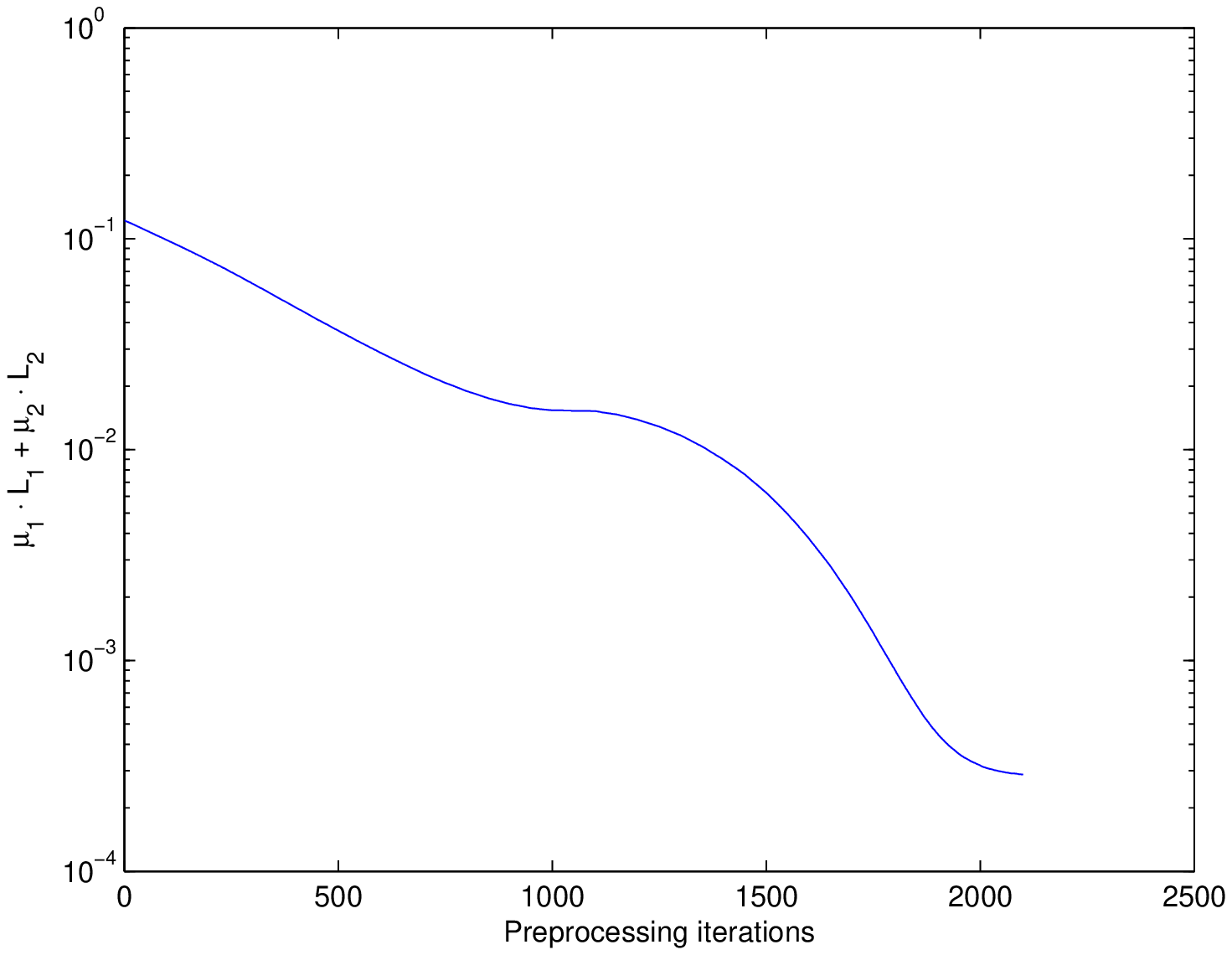}
 \caption{Evolution of the total functional $L^{\mathrm{(MF)}}$ as given by Eq.~(\ref{eq:LMF}) \emph{(top)}
and of the sum $\mu_1^{(\mathrm{MF})} {L_1}/{N_{L_1}} +  \mu_2^{(\mathrm{MF})} {L_2}/{N_{L_2}}$ of the normalized total force and the normalized total torque 
\emph{(bottom)} during the minimization in \emph{ppMF}
 on a logarithmic scale.
Both quantities are calculated for the enlarged,
$320\times 320$ pixel$^2$ magnetogram.}
 \label{fig:LMF}
\end{figure}

We argue that the reason for the differences in the capabilities to reduce the total magnetic force
and the total magnetic torque lies in the scheme of the simulated annealing used for the
minimization in \emph{ppMF}, compared to the Newton-Raphson scheme employed by \emph{ppTW}.
While \emph{ppTW} is forced to stop at the nearest local minimum it can find, \emph{ppMF} is able to leave this minimum and may find another one.
In Fig.~\ref{fig:LMF}, 
which shows the evolutions of the total functional $L^{\mathrm{(MF)}}$
and of the sum of the normalized total force and the normalized total torque in the course of the minimization in \emph{ppMF},
one can see that after roughly 1000 iterations a local minimum is found, but then this minimum
is left to find a deeper one after about 2000 iterations. The values of force and torque
in the first minimum are approximately the ones of \emph{ppTW}. So it seems possible that \emph{ppTW} was forced to stay at this minimum. An alternative possibility
is that \emph{ppTW} was able to find the global minimum, but that for the parameters
chosen, for instance as a result of stronger smoothing, the above minimum
is not the deepest one.
Furthermore, the pathways of the minimization in the two algorithms will in general be
different.

While for the enlarged, $320\times 320$ pixel$^2$ magnetogram, to which the preprocessing is applied,
the net flux per unit area pixel$^2$,
\begin{equation}
M_f = \frac{1}{\sum P} \cdot \frac{\sum_P B_z}{\sum_P \vert B_z \vert} \,,
\end{equation}
nearly vanishes  (\emph{unpreprocessed}: $2 \times 10^{-7}$, \emph{ppTW}: $3 \times 10^{-7}$, \emph{ppMF}: $4 \times 10^{-7}$), the vector magnetogram is not flux balanced to the same degree,
see third row in Table~\ref{tab:pp}. \emph{ppTW} preserves the measured net flux through the vector
magnetogram area, while \emph{ppMF} increases it slightly. For the enlarged, $320\times 320$ pixel$^2$ field
of view, both preprocessing routines leave the flux balanced. So \emph{ppMF} moderately redistributes the magnetic flux
between the area of the vector magnetogram and its surroundings.

\begin{figure*}
 \centering
 \includegraphics[width=0.3\textwidth]{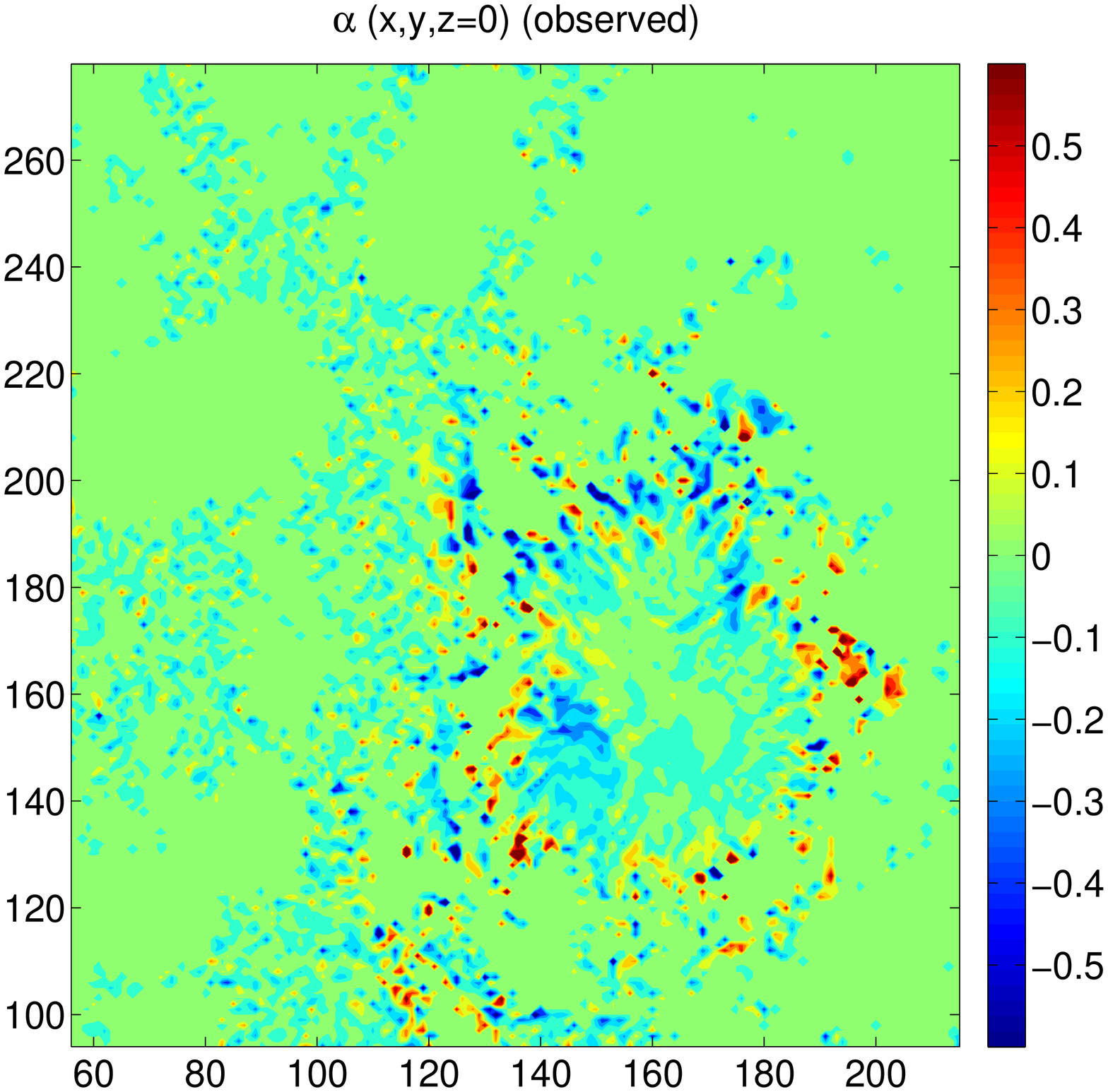}
 \includegraphics[width=0.3\textwidth]{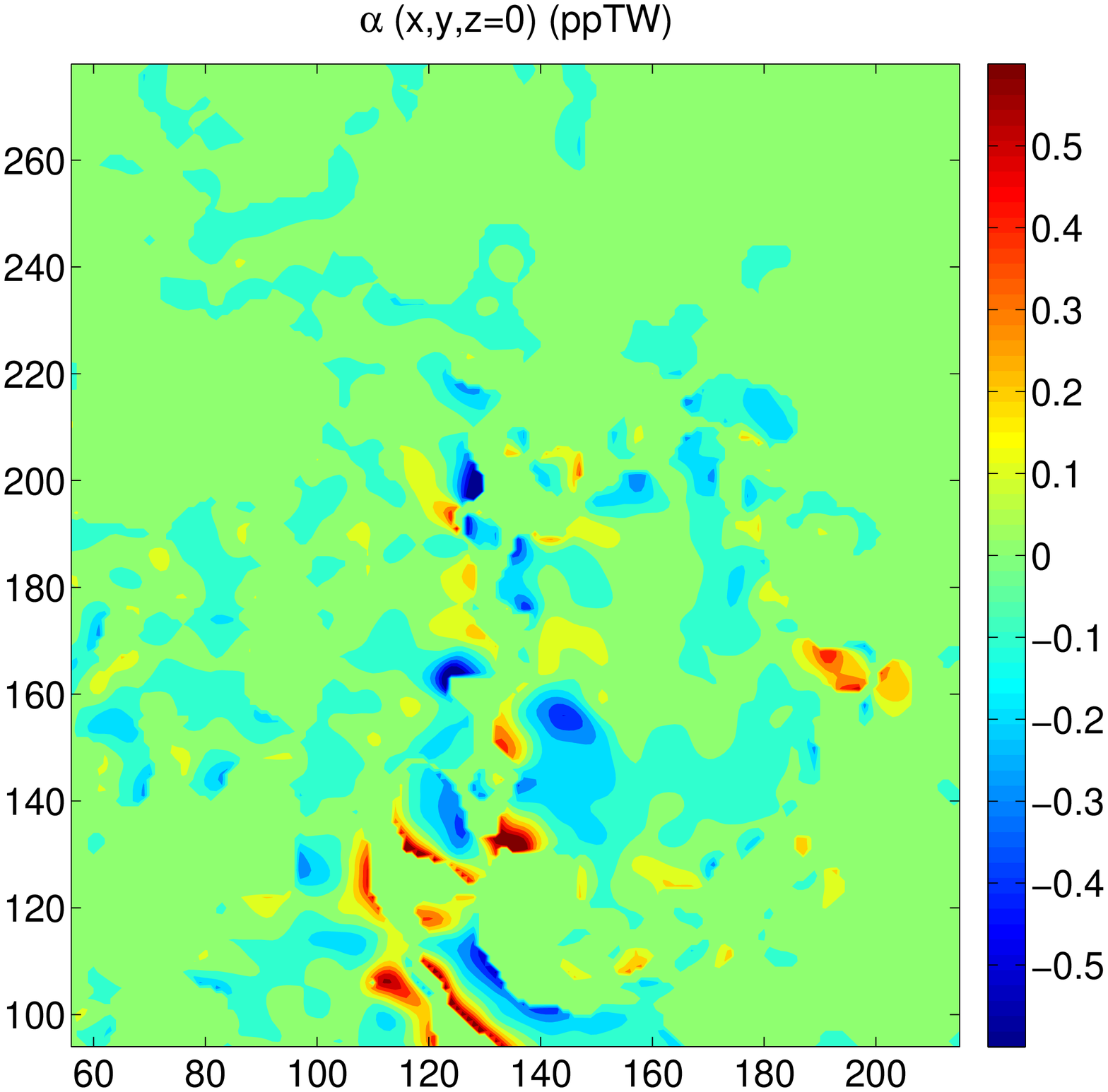}
 \includegraphics[width=0.3\textwidth]{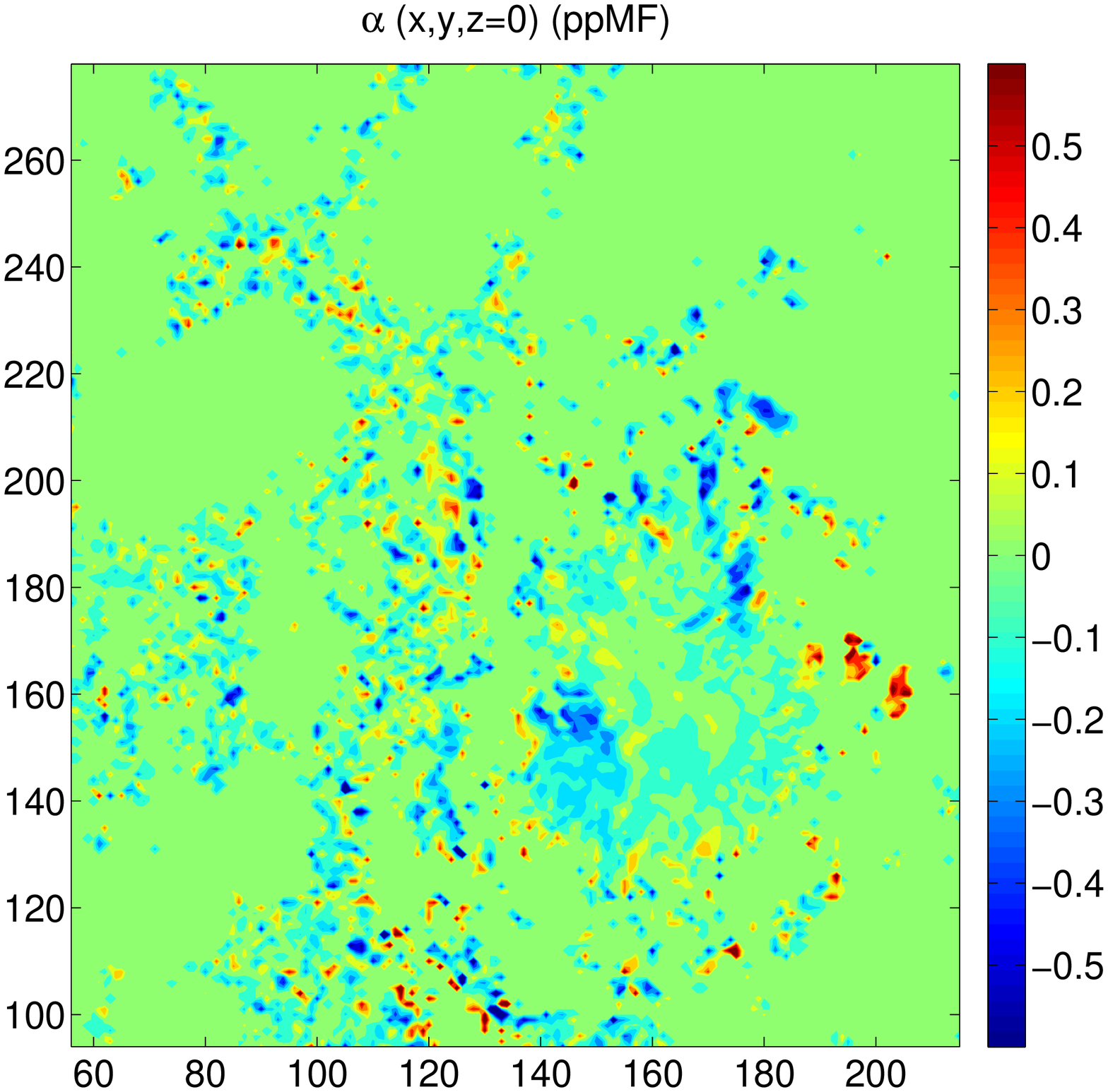}
  \caption{Contours of the function $\alpha(x,y,z=0)$. \emph{From left to right:} Unpreprocessed, \emph{ppTW}, and \emph{ppMF}.
The length unit is pixel ($580\,\mathrm{km}$) and $\alpha$ is measured in pixel$^{-1}$. }
\label{fig:alpha}
\end{figure*}

The fourth and fifth rows of Table~\ref{tab:pp},  and Fig.~\ref{fig:alpha}
show how the preprocessing routines change the function
\begin{equation}\label{alpha}
\alpha (x,y,z=0) = \frac{\partial_x B_y(x,y,z=0) - \partial_y B_x(x,y,z=0)}{B_z(x,y,z=0)}.
\end{equation}
Since this quantity depends on all three vector components, and is sensitive to noise
due to the differentiations needed to calculate it, it is expected to be particularly
strongly affected by the preprocessing.
The maximum absolute value of $\alpha$ is significantly decreased by both
procedures (by about 50\% by \emph{ppTW} and by about 35\% by \emph{ppMF}, see fourth row
in Table~\ref{tab:pp}). The mean value of $|\alpha|$ over the vector magnetogram area,
on the other hand,
is only weakly changed by both
procedures (fifth row in Table~\ref{tab:pp}). The unpreprocessed $\alpha$ function shown in Fig.~\ref{fig:alpha} (left) is highly complex and predominantly negative. \emph{ppTW} yields a much smoother $\alpha$ than does \emph{ppMF} (Fig.~\ref{fig:alpha}, middle and right), confirming the above observation that \emph{ppTW}  tends to remove small-scale structures much more than \emph{ppMF}.

\begin{figure*}
  \centering
\includegraphics[width=0.3\textwidth]{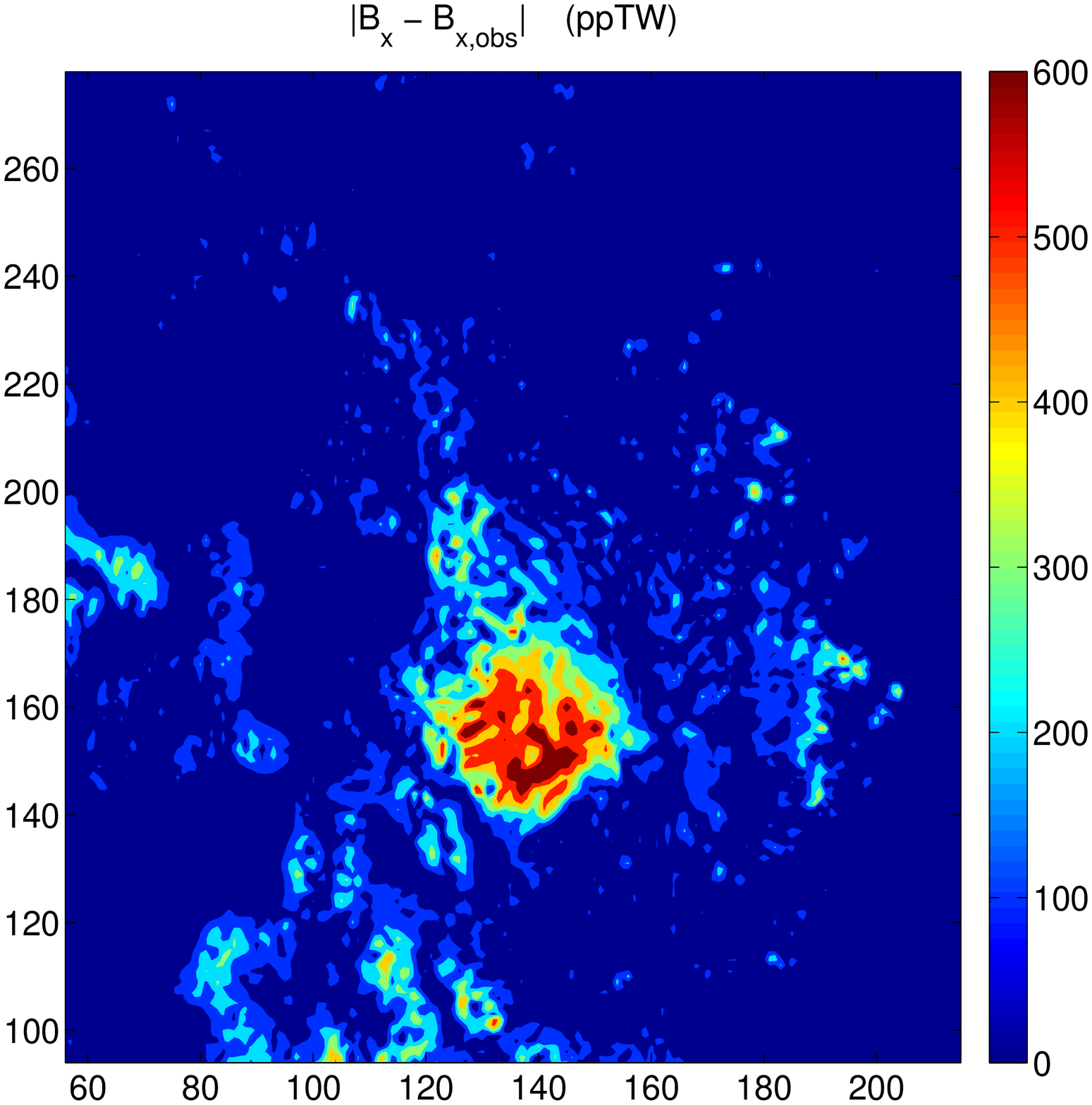}
\includegraphics[width=0.3\textwidth]{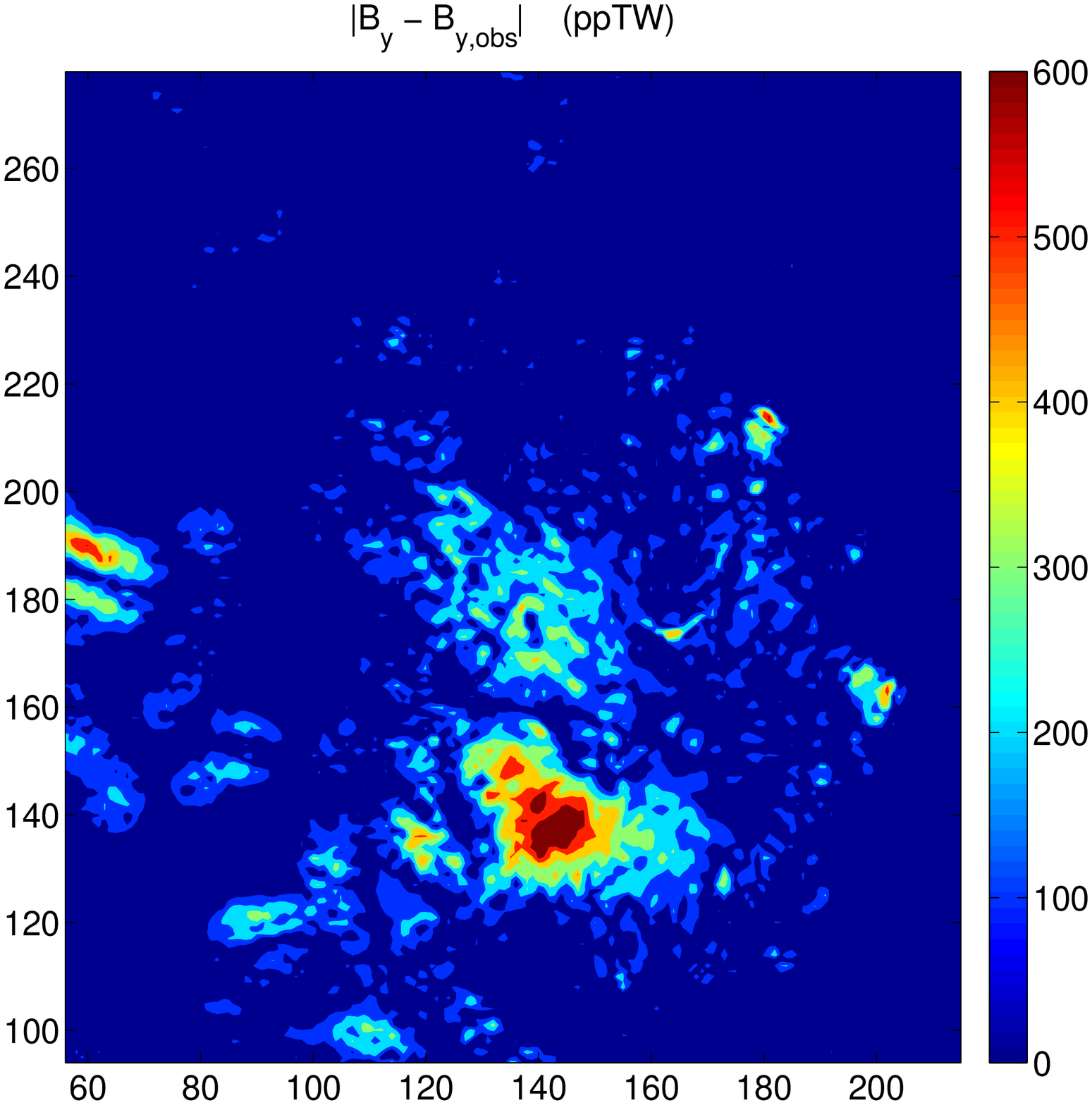}
\includegraphics[width=0.3\textwidth]{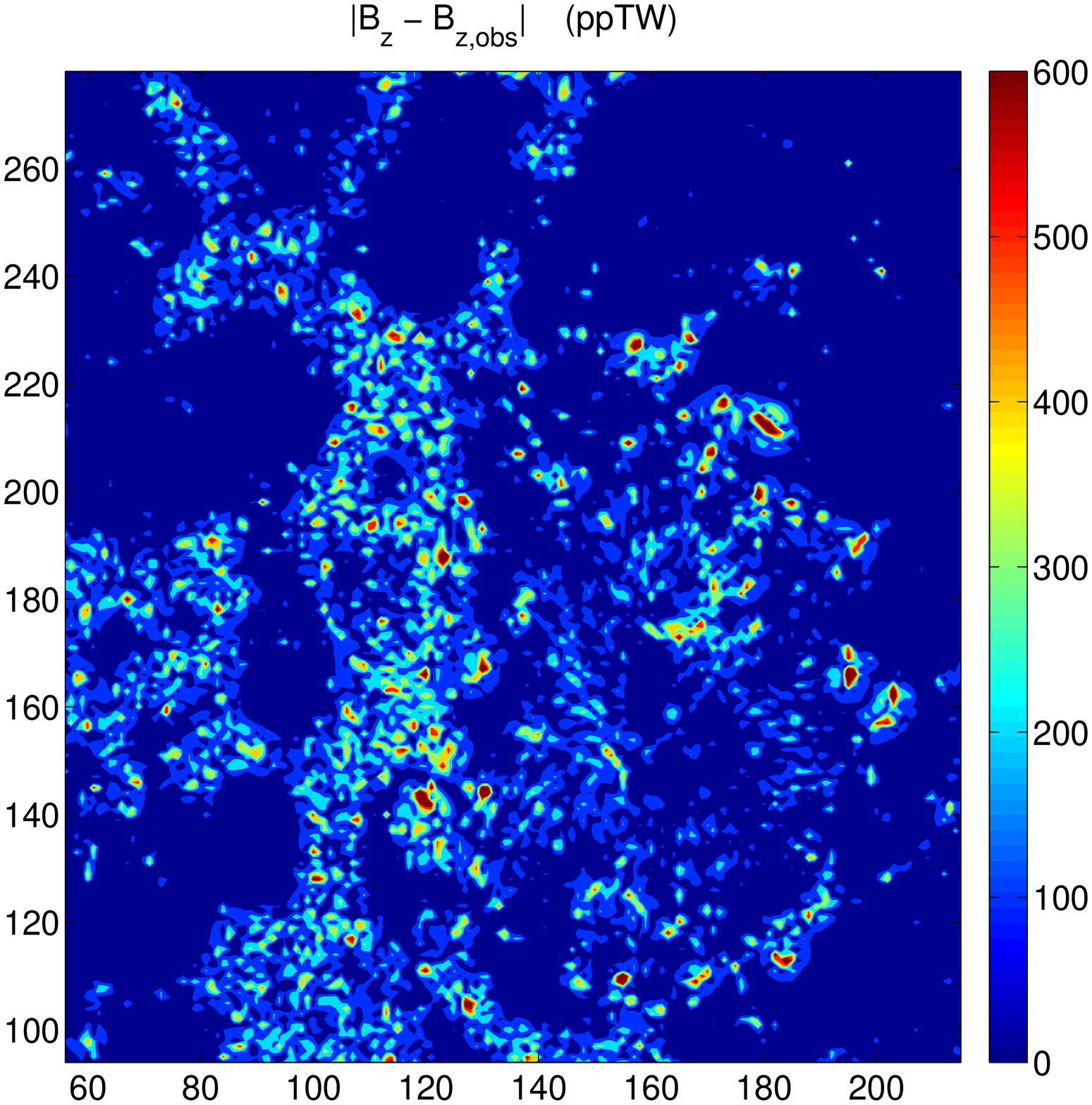}\\
\includegraphics[width=0.3\textwidth]{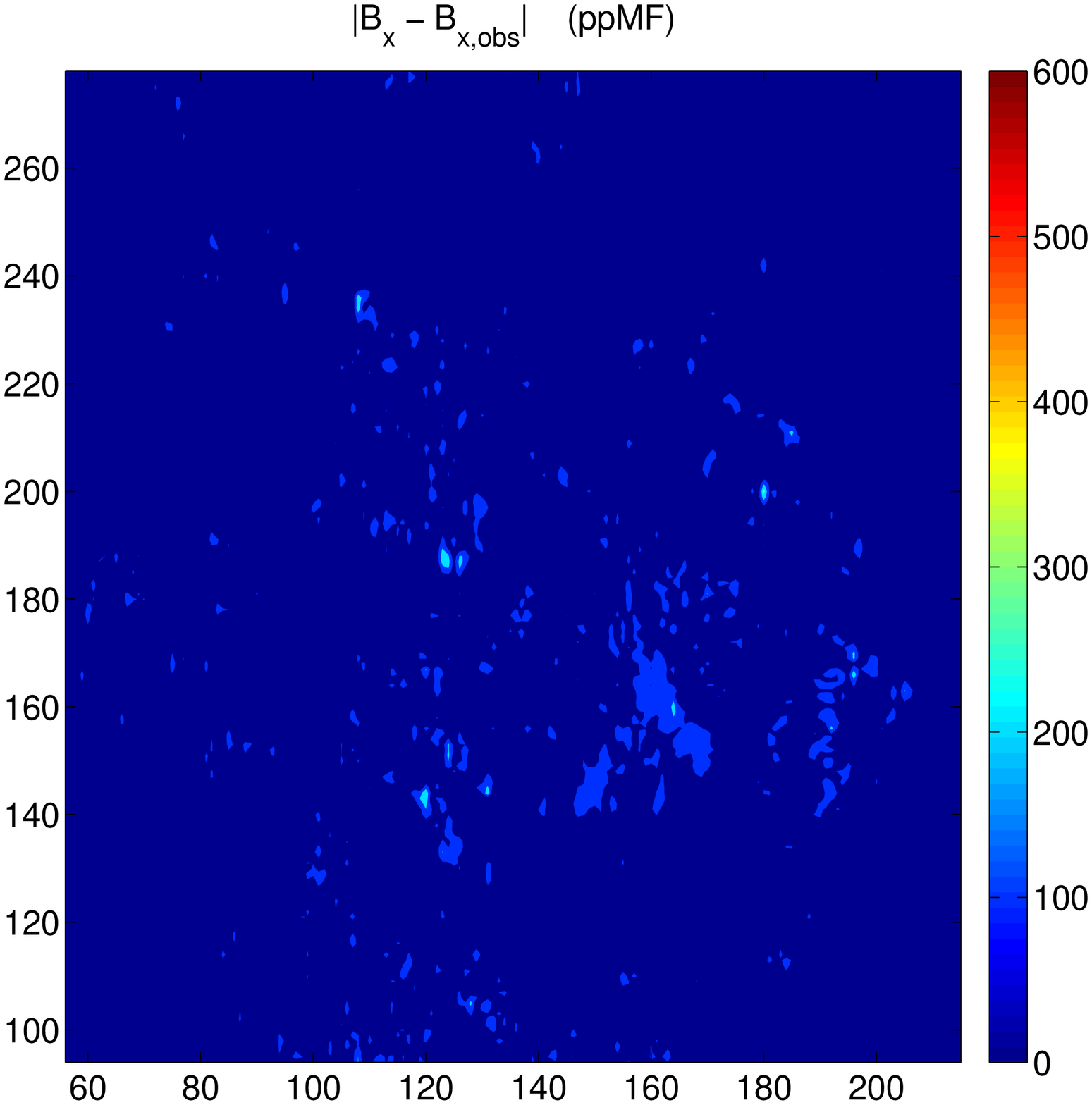}
\includegraphics[width=0.3\textwidth]{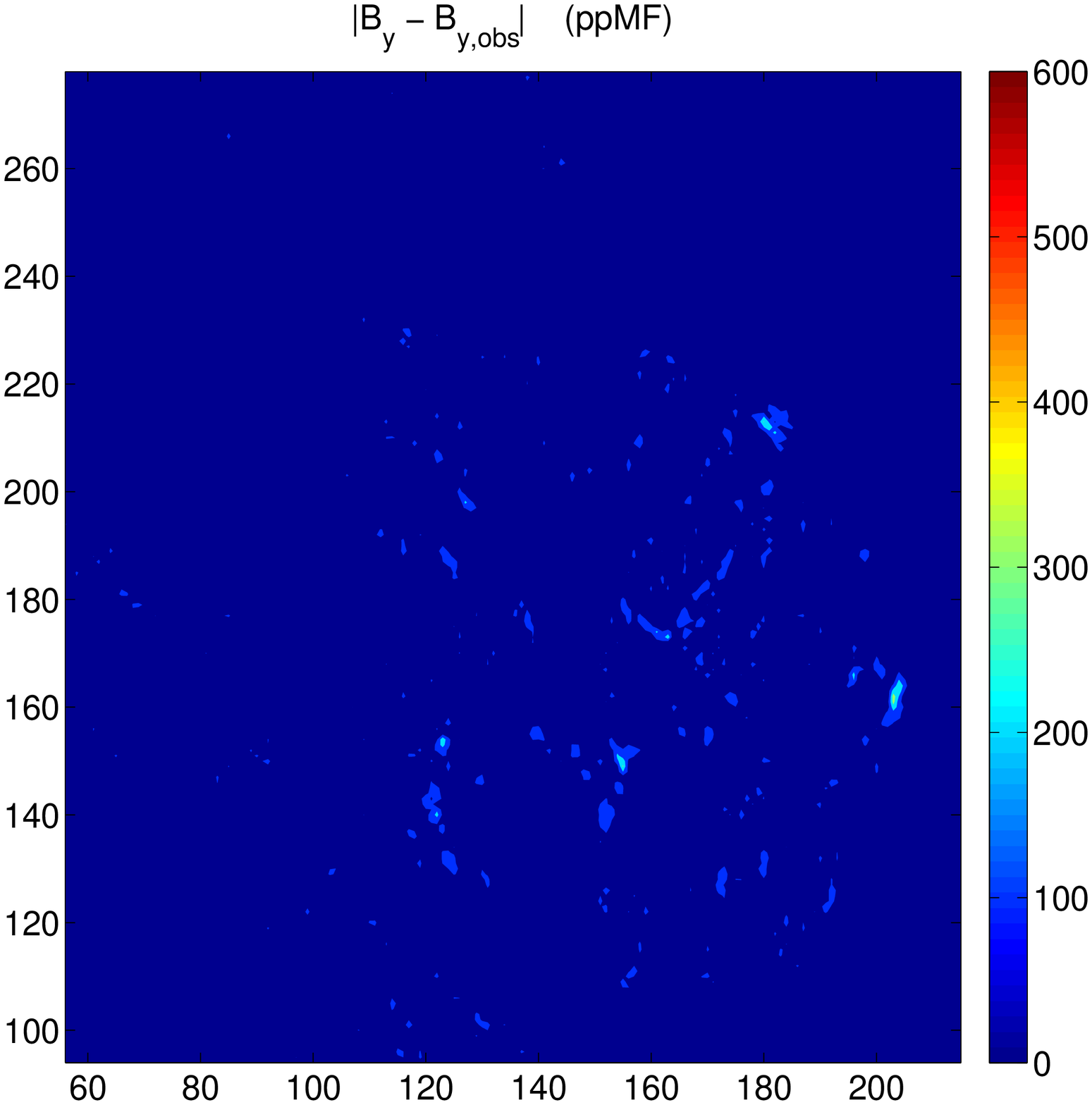}
\includegraphics[width=0.3\textwidth]{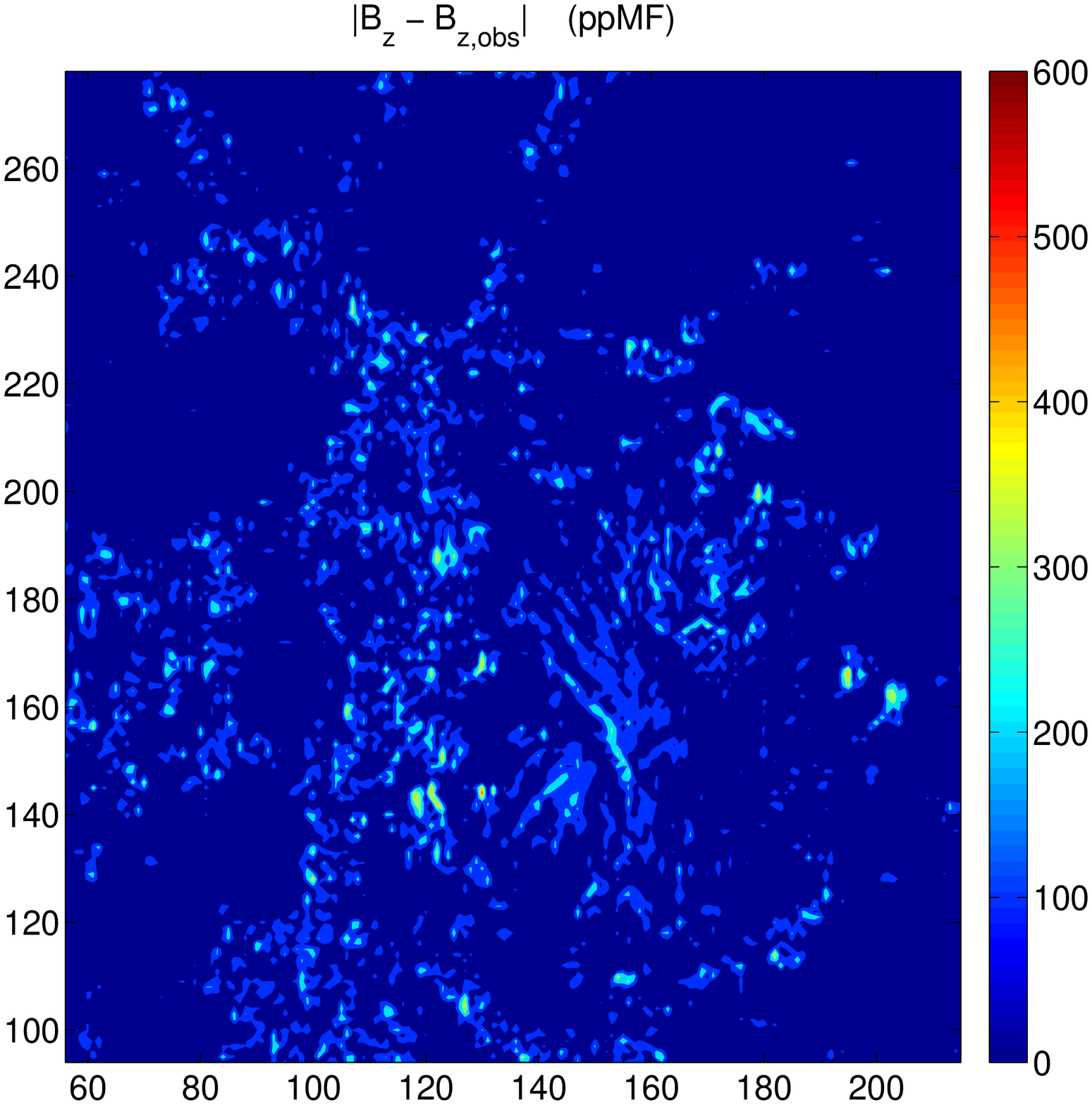}
\caption{Absolute values of the differences between the preprocessed and observed
magnetograms for \emph{ppTW} \emph{(top row)} and \emph{ppMF} \emph{(bottom row)}.
\emph{From left to right:} Differences for $B_{x}$, $B_{y}$, and $B_{z}$.
The magnetic field is measured in G and the length unit is pixel ($580\,\mathrm{km}$). }
  \label{fig:diffTW2UP}
\end{figure*}

Finally, Table~\ref{tab:diff10953}
\begin{table}
\caption{Absolute values of the differences between the preprocessed and observed fields.}
\centering
\begin{tabular}{ccc} 
 \hline\hline
Preprocessing method & \emph{ppTW} & \emph{ppMF}\\
 \hline
max. ($\vert B_x - B_{x,obs} \vert$) [G] & 306.9 & 119.9\\
max. ($\vert B_y - B_{y,obs} \vert$) [G] & 132.8 & 115.7\\
max. ($\vert B_z - B_{z,obs} \vert$) [G] & 215.0 & 116.6\\
$ \left< \vert B_x - B_{x,obs} \vert \right>$ [G] & 79.7 & 73.0\\
$\left< \vert B_y - B_{y,obs} \vert \right>$ [G] & 72.9 & 73.4\\
$\left< \vert B_z - B_{z,obs} \vert \right>$ [G] & 93.8 & 81.8\\
\hline
\end{tabular}
\label{tab:diff10953}
\end{table} 
shows the absolute values of the differences between the preprocessed and observed magnetograms,see also Fig.~\ref{fig:diffTW2UP}. 
\emph{ppTW} leads to great changes in strong field areas, but leaves weak field areas mostly unchanged. \emph{ppMF}, on the other hand, changes the whole magnetogram, but not as strongly as \emph{ppTW}. The average changes of $B_x$, $B_y$, and $B_z$  over the vector magnetogram are nearly identical for \emph{ppTW} and \emph{ppMF}.
The differences between the two methods may result from the fact that in \emph{ppTW} the deviation from
the original magnetogram is controlled globally via the subfunctional $L_3$,
while \emph{ppMF} uses a local control at each grid point.

\subsection{Effects of the preprocessing on the results of extrapolations}
\label{sec_effects_field}

The observed and the two preprocessed maps of the photospheric magnetic field vector
are now used as
input data for extrapolations using the optimization method of \citet{wie04}
and the magnetofrictional relaxation method of \citet{valklifuh07}.

The side and top boundaries are treated differently
by the two extrapolation  methods.
The  method of \citet{wie04} prescribes the magnetic vector as given by an initial
potential field. Our extrapolations with this method
are done, as the extrapolations in \citet{derosetal09} were done,
for a cuboid above the enlarged vector magnetogram, and with the boundary
condions at the side and top boundaries given by the values of the potential field calculated for the encompassing $512\times 512\times 512$ pixel$^3$ cube.
Contradictions between strong currents in the region above the vector magnetogram
and the assumption of a potential magnetic field at the side boundaries
can be mitigated by placing the side boundaries far away from the vector magnetogram area.

The method of \cite{valklifuh07}, as used here,
applies a special kind of 
open boundary conditions at the side and top boundaries,
where the normal field component is determined from inner field
values such as to ensure the solenoidal property of $\vec{B}$,
and the transverse field is obtained from
a fourth-order polynomial extrapolation of interior field values to the boundary;
this precedure works with a buffer layer of three grid points at the boundaries. 
These special boundary conditions allow extrapolations without assuming a potential field at the side and top boundaries.
Accordingly, our extrapolations using the method of \cite{valklifuh07} start from the smaller photospheric area of the vector magnetogram.
 That is to say, the solution domain for these extrapolations
is a cuboid with a height of 256 pixels above the measured vector magnetogram. 
The necessity of modelling the transverse photospheric field outside the
vector magnetogram area is avoided. Such modelling normally produces
unphysical current concentrations at and above the boundary of the vector magnetogram.
Indeed, using the \citeauthor{valklifuh07} method with an embedding in the larger
line-of-sight magnetogram \citep{derosetal09} leads to worse extrapolation results.

The potential field calculated for the encompassing
$512\times 512\times 512$ pixel$^3$ cube provides the intial field for both types of nonlinear force-free field computations.
The analysis of the extrapolation results
is always restricted to a cuboid above the area of the measured vector magnetogram,
still slightly reduced by the \citeauthor{valklifuh07} buffer layer.

For measuring the degree of force-freeness of the extrapolated fields, we use the metric \citep[cf.][]{whesturou00,schrietal06,valklifuh07,derosetal09}
\begin{align}\label{cwsin}
\begin{split}
 CW_{\sin} &= \frac{\sum_i \left| \vec J_i \right| \sigma_i}{\sum_i \left| \vec J_i \right|} \\
& \mbox{with}\;
\vec{J}_i=(\nabla\times\vec{B})_i,\;
\sigma_i = \frac{\left| \vec J_i \times \vec B_i \right|}{\left| \vec J_i \right| \left| \vec B_i \right|} = \left| \sin \theta_i \right|\,,
\end{split}
\end{align}
where the summation is over the grid points in the considered three-dimensional
domain $V$.
This is a current-weighted average of the sine of the angle $\theta$ between the
magnetic field $\vec{B}$ and the current density $\vec{J}$,
with $CW_{\sin}=0$ for an exactly force-free magnetic field.

Another point of interest is how well the codes fulfill the solenoidal condition, and how this is affected by the preprocessing. Here we use the metric
\begin{equation}
 \left< |\nabla\cdot\vec{B}|/|\vec{B}| \right> = \frac{1}{M} \sum_i\frac{\left|\nabla\cdot\vec{B}_i\right|}
{\left|\vec{B}_i\right|},
\end{equation}
with $M$ being the number of grid points in $V$.

Similarly, the relative magnetic energy in the volume, defined as
\begin{equation}
 \varepsilon = \frac{\sum_i \vec{B}_i^{2}}{\sum_i\vec{B}_{\mathrm{potential},i}^{2}}
\end{equation}
can be revealing. $\vec{B}_{\mathrm{potential}}$ is the potential field in $V$ whose normal component $B_n$
on the boundary $\partial V$ is identical to that of $\vec{B}$.
For a given distribution of $B_n$ on $\partial V$, $\vec{B}_{\mathrm{potential}}$ minimizes the magnetic
energy content of $V$, and $\varepsilon>1$ for any non-current-free magnetic field $\vec{B}$.
It has been observed, however, that nonlinear force-free extrapolations from observed magnetograms can produce 
fields with $\varepsilon<1$. It has also been noted that this pathological behavior can be corrected
by preprocessing the magnetograms \citep{metetal07_abstract}.

In Table~\ref{tab:ext},
\begin{table*}
\caption{Comparison of extrapolation results for different preprocessing and extrapolation methods.$^{\mathrm{a}}$}
\begin{center}
\begin{tabular}{c|cccccc}
 \hline\hline
Extrapolation  & Boundary map & $CW_{\sin}$ & $\left< \left| \vec{j} \times \vec{B} \right| \right> $ & $\mathrm{max}\,(\left| \vec{j} \times \vec{B} \right|) $ & $\left<|\nabla\cdot\vec{B}|/|\vec{B}|\right>$ &  $\varepsilon$\\
 \hline
Magneto- & unpreprocessed & 0.12 & 0.12 & 62.1 & 10.8 & 0.67\\
frictional & \emph{ppTW} &0.15 & 0.21 & 88.3  & 6.4 & 1.14\\
method & \emph{ppMF} & 0.08 & 0.14 & 62.2  & 3.6 & 1.12\\
\hline
Optimi- & unpreprocessed & 0.35 & 0.23 & 97.6  & 7.2 & 0.87\\
zation & \emph{ppTW} & 0.44 & 0.30 & 134.3 & 3.8 & 1.05\\
method& \emph{ppMF} & 0.29 & 0.20 & 98.9  & 1.2  & 1.02\\
\hline
\end{tabular}
\end{center} 
$^{\mathrm{a}}$
$CW_{\sin}$ and $\varepsilon$ are dimensionless,
 $\left< \left| \vec{j} \times \vec{B} \right| \right> $ and $\mathrm{max}\,(\left| \vec{j} \times \vec{B} \right|) $ are measured in $10^{3}\,\mathrm{G}^2\,\mathrm{pixel}^{-1}$, 
and $\left< |\nabla\cdot\vec{B}|/|\vec{B}| \right>$ is measured
in $10^{-7}\,\mathrm{pixel}^{-1}$
($1\,\mathrm{pixel}=580\,\mathrm{km}$).
\label{tab:ext}
\end{table*} 
the above metrics are given for the results of extrapolations using either the magnetofrictional
relaxation method or the optimization method and starting from the observed magnetogram or a magnetogram that was
preprocessed by \emph{ppTW} or \emph{ppMF}. It is seen in the table that for both extrapolation methods, only
\emph{ppMF} improves (i.e., makes smaller) the $CW_{\sin}$ values of the calculated fields. 
On the other hand, their $\left<|\nabla\cdot\vec{B}|/|\vec{B}|\right>$ values
are improved (i.e., brought closer to zero) for both extrapolation methods
by both \emph{ppTW} and \emph{ppMF}. Similarly, the relative magnetic energy,
$\varepsilon$, which is at unphysical values below 1 for the fields obtained
by the extrapolations starting from the observed magnetogram, is raised to values larger than 1 for both extrapolation methods by both \emph{ppTW} and \emph{ppMF}.
 Additionally, we examined the average and maximum values of the Lorentz force
 in the extrapolation volume. \emph{ppMF} leaves these values practically unchanged,
while \emph{ppTW} moderately increases them. This is largely in agreement with
the results for the $CW_{\mathrm{sin}}$ value, which stronger weights areas
with a high current density.

\begin{figure*}
 \centering
\includegraphics[width=0.3\textwidth]{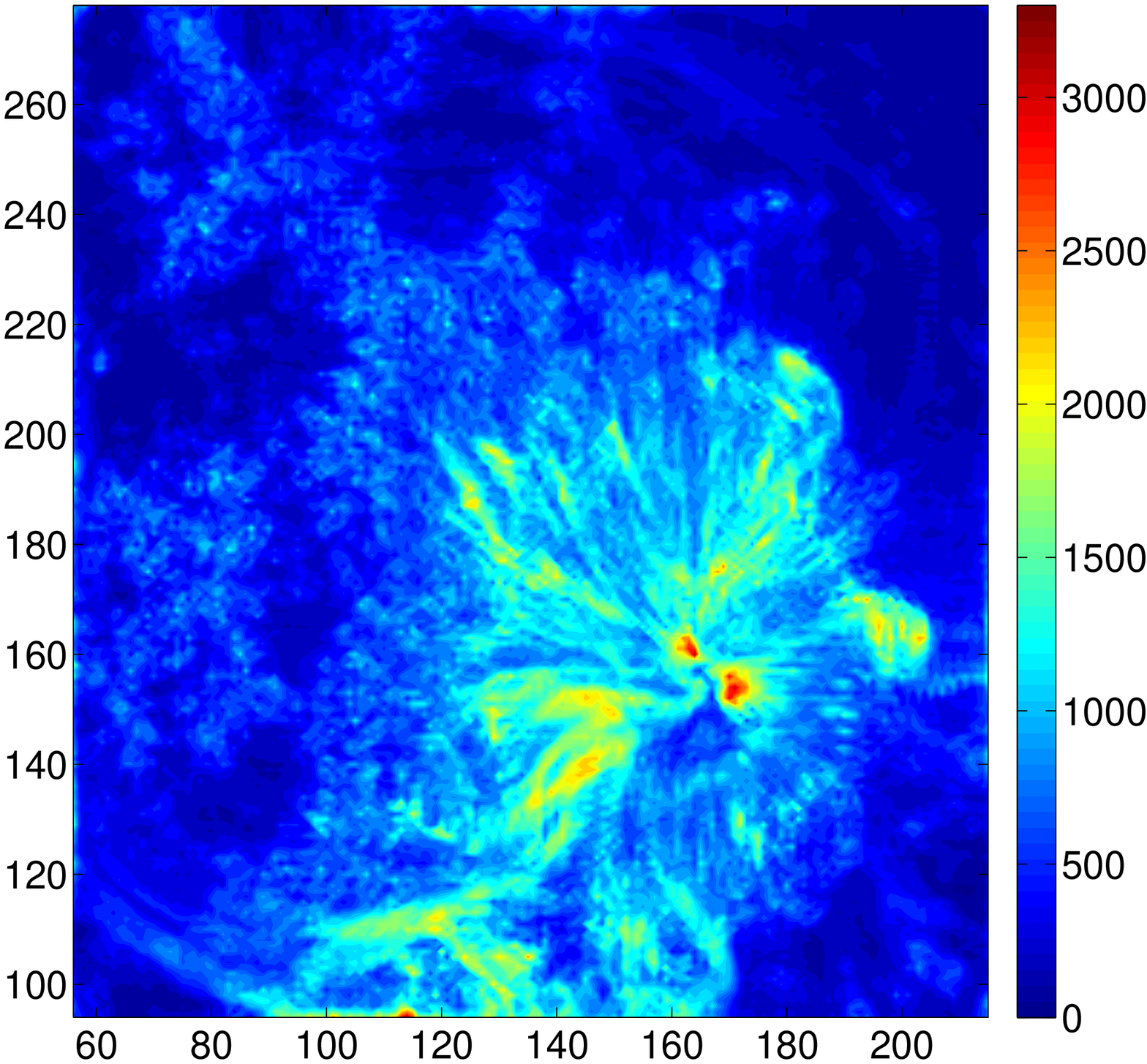}
\includegraphics[width=0.3\textwidth]{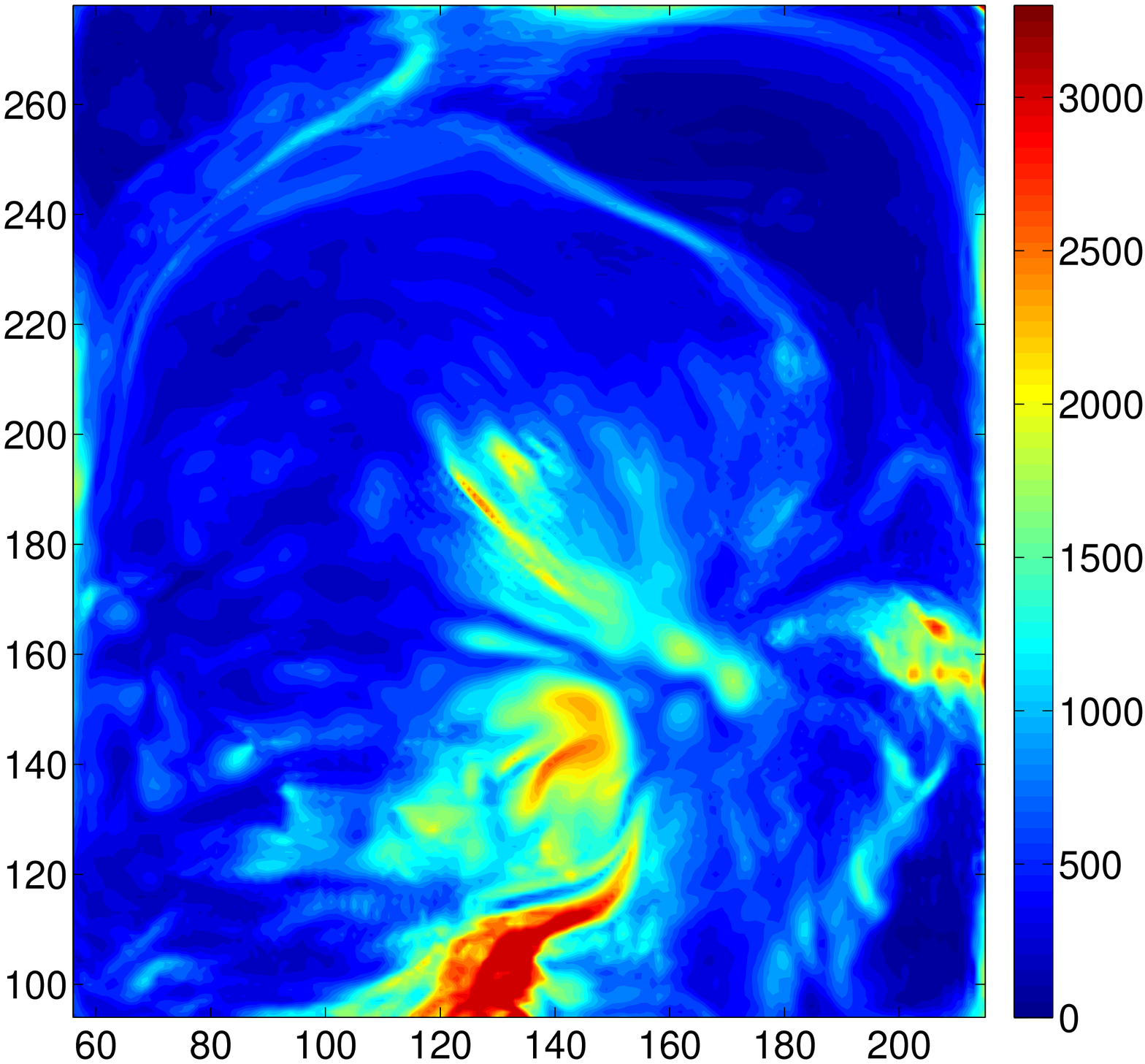}
\includegraphics[width=0.3\textwidth]{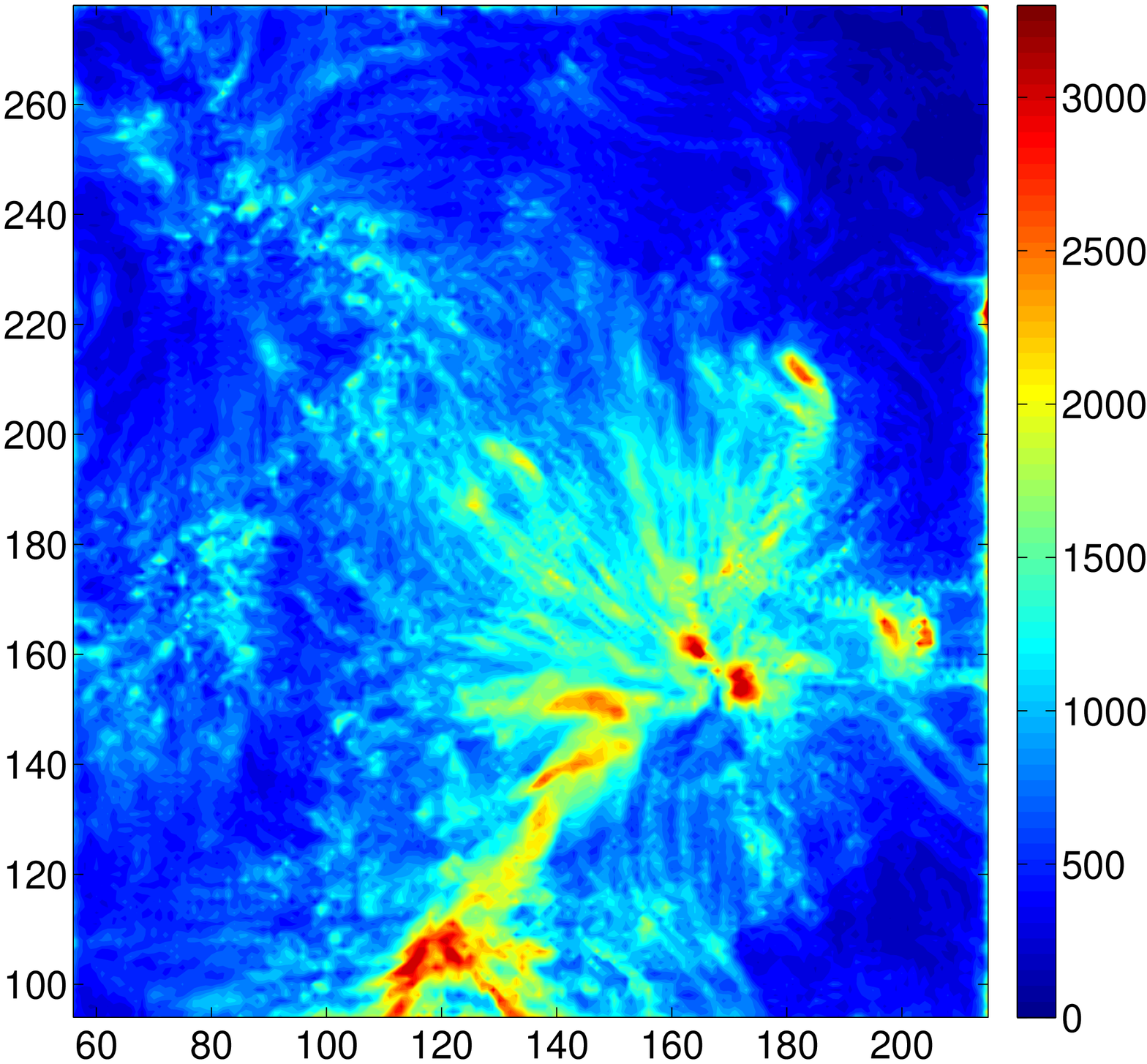}\\
\includegraphics[width=0.3\textwidth]{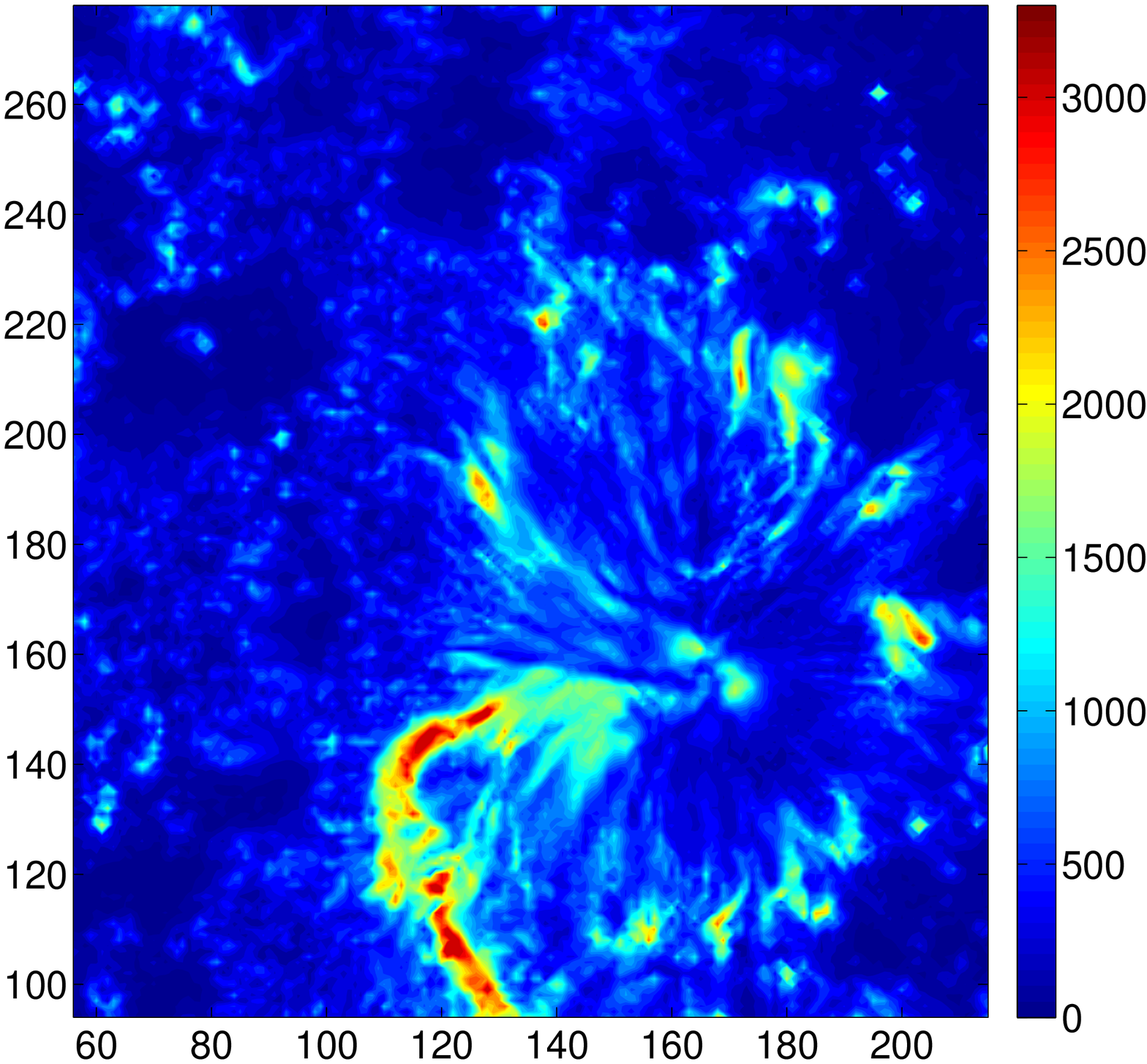}
\includegraphics[width=0.3\textwidth]{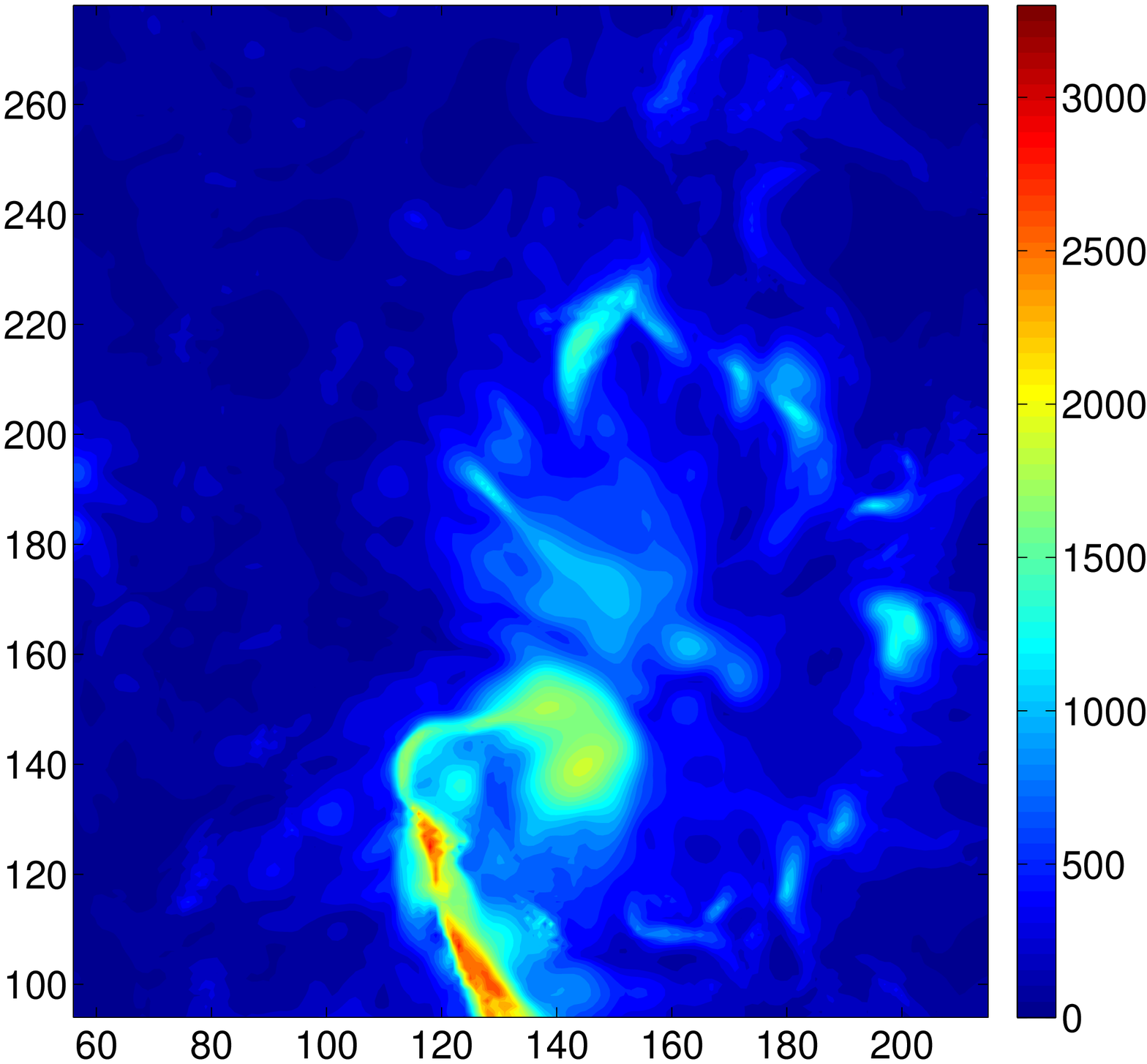}
\includegraphics[width=0.3\textwidth]{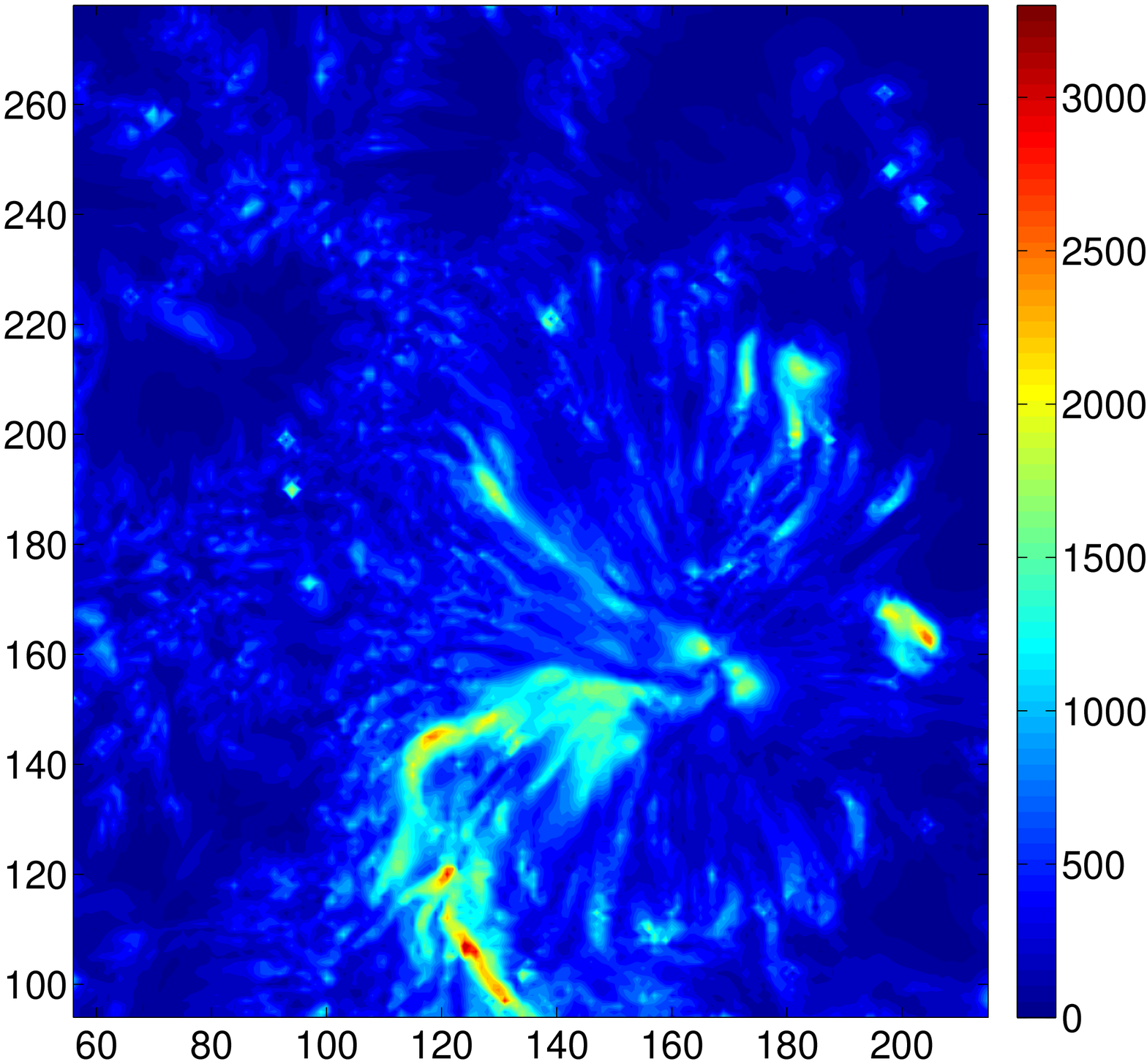}
 \caption{Absolute value of current density, $|\nabla\times\vec{B}|$, integrated vertically over a 12 pixel thick bottom layer, in G. \emph{Top row:} Extrapolation with magnetofrictional method. \emph{Bottom row:} Extrapolation with optimization method. \emph{From left to right:} Unpreprocessed, \emph{ppTW}, and \emph{ppMF}.
The length unit is pixel ($580\,\mathrm{km}$).} 
 \label{fig:normcurr}
\end{figure*}

Fig.~\ref{fig:normcurr},
finally, compares current densities for the six extrapolation/preprocessing cases in the form of
$|\nabla\times\vec{B}|$ integrated vertically over a 12 pixel thick bottom region.
For both extrapolations starting from the unpreprocessed vector magnetogram, the obtained current densities show
very complex patterns.

In the case of the magnetofrictional extrapolation method (see Fig.~\ref{fig:normcurr}, \emph{top row}), the
current density obtained when applying \emph{ppMF} appears to be very similar to that for the case
with no preprocessing. This means that  \emph{ppMF} here leads to a solution close to that obtained without
preprocessing, but with a physical magnetic energy content and in better agreement with the force-free 
and solenoidal conditions (cf. Table~\ref{tab:ext}). The field calculated by magnetofrictional extrapolation after
applying \emph{ppTW}, on the other hand, seems to be a much smoother and less structured version of the one calculated without preprocessing, at the same time being more physical in having an energy content in excess of the potential field value and in better fulfilling the solenoidal condition.

For the extrapolations using the optimization method (see Fig.~\ref{fig:normcurr}, \emph{bottom row}),
the two preprocessing algorithms seem to have broadly similar effects on the current density.
If one compares the cases with no preprocessing and with \emph{ppTW} applied, one can again see that \emph{ppTW}
smoothes the current density strongly. In the case of preprocessing with \emph{ppMF} the smoothing effect
is also observable, though to a lesser degree. This may have to do with the fact that,
independently of the kind of preprocessing applied, the current density distributions
obtained by extrapolations with the optimization method show weaker contrasts than those
obtained by magnetofrictional extrapolations. Comparing the figures in Table \ref{tab:ext}
from the point of view of the solutions' consistency, we can see that
the magnetofrictional extrapolations are closer to force-free (by a
factor of three in CWsin), while the optimization method led to more
solenoidal reconstructions. The optimization method is relatively little
affected by the energy in the small scales  of the
unpreprocessed magnetogram, while the magnetofrictional method attained
higher free energies with preprocessed magnetograms. (We do not extend further these observations since the comparison of different extrapolation
techniques is not our subject here).

\section{Conclusions} \label{sec_conclusions}

We have compared and discussed the two so far existing methods for the preprocessing of solar vector magnetograms, namely, those of \citet[][\emph{ppTW}]{wieinhsak06}
and \citet[][\emph{ppMF}]{fuhseeval07}.
These methods make the magnetograms more suitable for nonlinear force-free 
extrapolations into three-dimensional magnetic fields in the chromosphere and corona.
Both methods follow the same strategy, namely, to minimize a functional $L$ of the
photospheric field values such as to simultaneously make small the total magnetic
force and the total magnetic torque on the volume considered and the amount of small-scale noise in the photospheric boundary data.

The two methods differ in their ways of minimizing the functional $L$.
While \emph{ppTW} employs a Newton-Raphson scheme, 
\emph{ppMF} uses the method of simulated annealing.
Furthermore, the task of reducing the small-scale noise, or smoothing, is solved differently. \emph{ppTW} applies a minimization of the absolute value of
$\Delta \vec B(x,y,z=0)$, which corresponds to algebraically averaging the photospheric field, while in \emph{ppMF} the smoothing is reached by a windowed-median averaging.
Finally, the degree of deviation of the modified fields from the original photospheric field during the minimization is controlled in different ways by the two methods. In \emph{ppTW}
a special subfunctional that measures the deviation is included as an additional
minimization target in $L$. This method of control is integral in character.
\emph{ppMF}, on the other hand, uses a local control at each grid point
in the form of an interval within which the field values have to stay.

The two preprocessing methods were applied to a vector magnetogram
of the recently observed active region NOAA AR 10953, which  had already been the target
in a comparative study of different extrapolation methods for nonlinear force-free
magnetic fields reported in \citet{derosetal09}.
Both preprocessing methods managed to significantly decrease the magnetic forces and
magnetic torques that act through the magnetogram area and that can cause incompatibilities with the assumption of force-freeness in the solution domain. The force and torque
decrease was stronger for \emph{ppMF} than for \emph{ppTW}.

Both methods also reduced the amount of small-scale
irregularities in the observed photospheric field,
where \emph{ppTW} led to a markedly smoother magnetogram than \emph{ppMF},
in accordance with the aim of \emph{ppTW} to mimic the expansion of the solar magnetic field between photosphere and chromosphere.
The average deviations of the preprocessed magnetograms from the observed magnetogram
are nearly identical for \emph{ppTW} and \emph{ppMF}, where \emph{ppTW} led to
greater changes in strong-field areas, leaving weak-field areas mostly unchanged,
while \emph{ppMF} weakly changed the whole magnetogram, thereby better preserving
patterns present in the original magnetogram. Similarly, for the function
$\alpha(x,y,z=0)$ \emph{ppTW} yielded a much smoother distribution than \emph{ppMF}.

The original magnetogram and the two preprocessed magnetograms were used as input data
for nonlinear force-free field extrapolations by means of
two different methods, namely the magnetofrictional relaxation method of \citet{valklifuh07} and the optimization method of \citet{wie04}.
Both \emph{ppTW} and \emph{ppMF}  corrected a pathological property of the fields
calculated by the extrapolations from the observed magnetogram,
namely they raised the magnetic energy content of the extrapolated fields
from values below to values above that
for the potential field with the same normal component on the boundary.
Also, the fields calculated from the preprocessed magnetograms fulfill the solenoidal condition for $\vec{B}$ better than those calculated without preprocessing.
Finally, some effects of the preprocessing, as those on the current density in a bottom
layer, seemed to be influenced by the the specifics
of the employed extrapolation method.

\begin{acknowledgements}
Hinode is a Japanese mission developed and launched by ISAS/JAXA, with NAOJ as domestic partner and NASA and STFC (UK) as international partners. It is operated by these agencies in co-operation with ESA and NSC (Norway). M. Fuhrmann was supported by DFG grant SE 662/11-1 and T. Wiegelmann by DLR grant 50 OC 0501.
\end{acknowledgements}

\end{document}